\documentclass[11pt]{article}

\usepackage{amsmath,amssymb}
\usepackage{array}
\usepackage{epsfig}

\usepackage{amsmath}
\usepackage{graphicx}
\usepackage{latexsym}
\usepackage{overcite}

\newcount\figureno     \figureno=0
\newdimen\figdim       \figdim=70mm
\def\figureinc{%
   \global\advance\figureno by 1%
}
\def\figcaption#1#2#3{\hbox to #2{\hss{\vbox{\hsize=#2 \parindent=0pt
        {\bf Figure \number\figureno#3 :\ }#1}}\hss}
}

\makeatletter
\renewcommand{\@biblabel}[1]{$^{#1}$}
\makeatother

\begin{document}
\baselineskip 100pt

{\large
\parskip.2in
\newcommand{\be}{\begin{equation}}
\newcommand{\ee}{\end{equation}}
\newcommand{\ben}{\begin{equation*}}
\newcommand{\een}{\end{equation*}}
\newcommand{\br}{\bar}
\newcommand{\fr}{\frac}
\newcommand{\lm}{\lambda}
\newcommand{\ra}{\rightarrow}
\newcommand{\al}{\alpha}
\newcommand{\bt}{\beta}
\newcommand{\z}{\zeta}
\newcommand{\pa}{\partial}
\newcommand{\hs}{\hspace{5mm}}
\newcommand{\up}{\upsilon}
\newcommand{\dg}{\dagger}
\newcommand{\sdil}{\ensuremath{\rlap{\raisebox{.15ex}{$\mskip
6.5mu\scriptstyle+ $}}\subset}}
\newcommand{\sdir}{\ensuremath{\rlap{\raisebox{.15ex}{$\mskip
6.5mu\scriptstyle+ $}}\supset}}
\newcommand{\vphi}{\vec{\varphi}}
\newcommand{\ve}{\varepsilon}
\newcommand{\acc}{\\[3mm]}
\newcommand{\dl}{\delta}
\def\tablecap#1{\vskip 3mm \centerline{#1}\vskip 5mm}
\def\p#1{\partial_#1}
\newcommand{\pd}[2]{\frac{\partial #1}{\partial #2}}
\newcommand{\pdn}[3]{\frac{\partial #1^{#3}}{\partial #2^{#3}}}
\def\DP#1#2{D_{#1}\varphi^{#2}}
\def\dP#1#2{\partial_{#1}\varphi^{#2}}
\def\xh{\hat x}
\newcommand{\Ref}[1]{(\ref{#1})}

\def\mod#1{ \vert #1 \vert }
\def\chapter#1{\hbox{Introduction.}}
\def\Sin{\hbox{sin}}
\def\Cos{\hbox{cos}}
\def\Exp{\hbox{exp}}
\def\Ln{\hbox{ln}}
\def\Tan{\hbox{tan}}
\def\Cot{\hbox{cot}}
\def\Sinh{\hbox{sinh}}
\def\Cosh{\hbox{cosh}}
\def\Tanh{\hbox{tanh}}
\def\Asin{\hbox{asin}}
\def\Acos{\hbox{acos}}
\def\Atan{\hbox{atan}}
\def\Asinh{\hbox{asinh}}
\def\Acosh{\hbox{acosh}}
\def\Atanh{\hbox{atanh}}
\def\frac#1#2{{\textstyle{#1\over #2}}}

\newcommand{\ie}{{\it i.e.}}
\newcommand{\cmod}[1]{ \vert #1 \vert ^2 }
\newcommand{\cmodn}[2]{ \vert #1 \vert ^{#2} }
\newcommand{\nhat}{\mbox{\boldmath$\hat n$}}
\nopagebreak[3]
\bigskip

\title{ \bf Supersymmetric formulation of polytropic gas dynamics and its invariant solutions }
\vskip 1cm

\bigskip
\author{
A.~M. Grundland\thanks{email address: grundlan@crm.umontreal.ca}
\\
Centre de Recherches Math{\'e}matiques, Universit{\'e} de Montr{\'e}al,\\
C. P. 6128, Succ.\ Centre-ville, Montr{\'e}al, (QC) H3C 3J7,
Canada\\ Universit\'{e} du Qu\'{e}bec, Trois-Rivi\`{e}res, CP500 (QC) G9A 5H7, Canada \acc A. J. Hariton\thanks{email address: hariton@crm.umontreal.ca}
\\
Centre de Recherches Math{\'e}matiques, Universit{\'e} de Montr{\'e}al, \\
C. P. 6128, Succ.\ Centre-ville, Montr{\'e}al, (QC) H3C 3J7, Canada \\} \date{}

\maketitle

\begin{abstract}

In this paper, a supersymmetric extension of the polytropic gas dynamics equations is constructed through the use of a superspace involving two independent fermionic variables and two bosonic superfields. A superalgebra of symmetries of the proposed extended model is determined and a systematic classification of the one-dimensional subalgebras of this superalgebra is performed. Through the use of the symmetry reduction method, a number of invariant solutions of the supersymmetric polytropic gas dynamics equations are found. Several types of solutions are obtained, including algebraic-type solutions and propagation waves (simple and double waves). Many of the obtained solutions involve arbitrary functions of one or two bosonic or fermionic variables. In the case where the arbitrary functions involve only the independent fermionic variables, the solutions are expressed in terms of Taylor expansions.

\end{abstract}

Byline: Supersymmetric polytropic gas and its solutions

PACS: 02.20.Sv, 12.60.Jv, 02.30.Jr

Keywords: supersymmetric model, Lie superalgebra, polytropic gas dynamics, symmetry reduction, invariant solutions

\newpage

\section{Introduction}

Symmetries of fluid dynamics have recently attracted a considerable amount of interest among physicists and mathematicians. Dispersionless hydrodynamic systems can be studied from a variety of points of view, of which the best understood are $(1+1)$-dimensional systems. Such systems include, among others, polytropic gases, the Chaplygin gas, the Born-Infeld equations and hydrodynamic systems written in terms of Riemann invariants. Such systems also include strings and Nambu-Goto membranes (see e.g. R. Jackiw \cite{Jackiw} and references therein).

The purpose of this paper is to construct a supersymmetric version of the polytropic gas dynamics equations in $(1+1)$ dimensions
\begin{equation}
\begin{split}
&\rho_t+\rho_xw+\rho w_x=0\\
& \\
&w_t+ww_x+A\rho^{\gamma}\rho_x=0,
\end{split}
\label{a1}
\end{equation}
where $w$ is the velocity of the fluid, $\rho$ is its density and $\gamma$ is the polytropic exponent. These equations are linked with the ones given in Das and Popowicz \cite{Das} (p. 3, equation (1))
\begin{equation}
{\partial v\over \partial t}=(vu)_x,\qquad {\partial u\over \partial t}=\left({u^2\over 2}+{v^{\gamma-1}\over (\gamma-1)}\right)_x
\label{Daseqtype1}
\end{equation}
(given in the form of conservation laws) through the transformation $w=-u$,\ \ $\rho=v$.
For the case where the polytropic exponent is $\gamma=-1$, we get the Chaplygin gas. Supersymmetric versions of the Chaplygin gas in $(1+1)$ and $(2+1)$ dimensions were formulated through the parametrization of the action for a superstring and supermembrane respectively \cite{Jackiw,Bergner,Polychronakos}.
The case $\gamma=2$ refers to the dispersionless limit of two bosonic equations \cite{Roz}. When $\gamma=3$, the equations represent the dispersionless limit of two non-interacting Korteweg-de Vries equations (known as Riemann invariants equations) \cite{Jackiw,Hariton9,Hariton10}. Finally, the $\gamma=4$ case is the dispersionless limit of a Boussinesq equation \cite{Ablowitz,Rogers}.

In the context of hydrodynamic-type equations, other models have been supersymmetrized and analyzed from the group-theoretical point of view. Examples of such models include the Korteweg-de Vries equation \cite{Mathieu,Labelle}, the Kadomtsev-Petviashvili equation \cite{Manin}, the scalar Born-Infeld model \cite{Hariton4}, a Gaussian fluid flow \cite{Hariton8} and a hydrodynamic system expressed in terms of Riemann invariants \cite{Fatyga,Hariton9,Hariton10}.
The results are of interest since we can construct certain classes of solutions with the freedom of arbitrary functions of one or two arguments involving bosonic and fermionic variables.

In the case of several integrable systems, the Lax pairs and multisolitonic solutions have been found for their supersymmetric extensions (see e.g. \cite{Grammaticos,Siddiq1,Siddiq2,Chaichian,Liu3}). It is natural in this context to consider other types of wave superpositions. It has been established that classical hydrodynamic-type systems admit simple and multiple Riemann wave (scattering and non-scattering) solutions \cite{Roz,Peradzynski}. The question arises as to whether or not the same holds true for a supersymmetric extension of such a model. We will answer this question in the case of the polytropic gas dynamics equations (\ref{a1}) by performing a symmetry analysis.

The supersymmetric extension that we construct in this paper differs from that of Das and Popowicz \cite{Das} in the sense that our superspace includes two independent fermionic variables instead of one. Also, the superfields which we propose are bosonic-valued instead of fermionic-valued. This is a different form of supersymmetric extension for the equations (\ref{a1}) which, to our knowledge, has not previously been considered. We perform a systematic group-theoretical analysis of our constructed supersymmetric model.

This paper is organized as follows. In section 2, we construct a supersymmetric extension of the polytropic gas dynamics equations (\ref{a1}) through a superspace and superfield formalism involving two independent fermionic variables. In section 3, we present in detail a number of symmetries of our supersymmetric system. For this purpose, we use a generalization of the method of prolongation of vector fields, extended so as to include both even and odd Grassmannian variables. The symmetry criterion is adapted to this case, and the superalgebra associated with the supersymmetric system is determined. We classify the one-dimensional subalgebras of this superalgebra into conjugation classes under the action of the associated supergroup, and this allows us to perform symmetry reductions in a systematic way. In section 4, we proceed to find the invariants and reduced systems corresponding to those subalgebras with standard invariants structures. The other subalgebras, possessing a nonstandard invariant structure, are discussed separately. We propose a new method for solving the reduced equation. This leads to solutions of the supersymmetric polytropic gas model, many of which contain the freedom of arbitrary functions of one or two arguments which are expressed in terms of bosonic or fermionic variables. In the case when the arbitrary functions involve only the independent fermionic variables, we were able to express the solution in terms of a naturally truncated Taylor expansion. Finally, section 5 contains the final remarks and future outlook.

\section{Supersymmetric extension}

The supersymmetric extension is constructed by first considering the following superspace and superfield formalism. We extend the space of independent variables $\{(x,t)\}$, which presently includes only bosonic variables, to the superspace $\{(x,t,\theta_1,\theta_2)\}$ which also includes the independent fermionic variables $\theta_1$ and $\theta_2$. The variables $x$ and $t$ represent the bosonic (even Grassmannian) coordinates on 2-dimensional Minkowski space, while the quantities $\theta_1$ and $\theta_2$ are anticommuting fermionic (odd Grassmannian) variables. We replace the real bosonic-valued fields of velocity $w(x,t)$ and density $\rho(x,t)$  by the bosonic superfields
defined as
\begin{equation}
\begin{split}
&W\left(x,t,\theta_1,\theta_2\right)=w(x,t)+\theta_1\phi_1(x,t)+\theta_2\phi_2(x,t)+\theta_1\theta_2F(x,t),\\
& \\
&P\left(x,t,\theta_1,\theta_2\right)=\rho(x,t)+\theta_1\psi_1(x,t)+\theta_2\psi_2(x,t)+\theta_1\theta_2G(x,t)
\end{split}
\label{b1}
\end{equation}
respectively, where $\phi_1$, $\phi_2$, $\psi_1$ and $\psi_2$ are fermionic fields and $F$ and $G$ are bosonic fields. The supersymmetric extension of the polytropic gas equations (\ref{a1}) is constructed in such a way that it is invariant under the supersymmetry transformations
\begin{equation}
x\rightarrow x-\underline{\eta}_1\theta_1,\quad \theta_1\rightarrow\theta_1+\underline{\eta}_1\qquad\mbox{ and }\qquad t\rightarrow t-\underline{\eta}_2\theta_2,\quad \theta_2\rightarrow\theta_2+\underline{\eta}_2,
\label{b2}
\end{equation}
where $\underline{\eta_1}$ and $\underline{\eta_2}$ are fermionic constants. In what follows, we use the convention that underlined constants represent fermionic parameters. These transformations are generated by the infinitesimal supersymmetry generators
\begin{equation}
Q_1=\partial_{\theta_1}-\theta_1\partial_x\qquad\mbox{ and }\qquad Q_2=\partial_{\theta_2}-\theta_2\partial_t,
\label{b3}
\end{equation}
which satisfy the anticommutation relations
\begin{equation}
\{Q_1,Q_1\}=-2\partial_x,\qquad \{Q_2,Q_2\}=-2\partial_t.
\label{zorrano}
\end{equation}
Here, we use the concept of supercommutation in the sense that we take the commutator of either two bosonic quantities or of a bosonic quantity and a fermionic quantity, but the anticommutator of two fermionic quantities.
In order to make our superfield theory manifestly invariant under the action of the supersymmetry generators $Q_1$ and $Q_2$, we write the supersymmetric system in terms of the covariant derivative operators
\begin{equation}
D_1=\partial_{\theta_1}+\theta_1\partial_x\qquad\mbox{ and }\qquad D_2=\partial_{\theta_2}+\theta_2\partial_t,
\label{b4}
\end{equation}
which possess the property that they anticommute with the supersymmetry generators $Q_1$ and $Q_2$
\begin{equation}
\begin{split}
\{D_1,D_1\}=&2\partial_x,\quad \{D_2,D_2\}=2\partial_t,\\ \{D_1,D_2\}=\{D_1,Q_1\}=&\{D_1,Q_2\}=\{D_2,Q_1\}=\{D_2,Q_2\}=0.
\end{split}
\label{b5}
\end{equation}
We supersymmetrize the polytropic gas dynamics equations (\ref{a1}) directly at the equation of motion level instead of from its Lagrangian form.
In order to build the supersymmetric extension, we evaluate the covariant derivatives of the superfields $W$ and $P$ of various orders. The most general supersymmetric extension of the polytropic gas equation (\ref{a1}) is constructed by considering linear combinations of the products of the various covariant derivatives of the superfields $W$ and $P$. These multiply together to produce the given terms as coefficients whose component reproduces each term of the classical equations (\ref{a1}). The result of this analysis gives the following form of the supersymmetric polytropic gas dynamics equations
\begin{equation}
\begin{split}
&P_t+a(D_1D_2P)_xW+b(D_2P)_x(D_1W)+c(D_1P)_x(D_2W)\\&+(c-b-a-1)P_x(D_1D_2W)+dP(D_1D_2W)_x+e(D_1P)(D_2W)_x\\&+f(D_2P)(D_1W)_x+(e-f-d-1)(D_1D_2P)W_x=0
\end{split}
\label{b6-epsilon}
\end{equation}
and
\begin{equation}
\begin{split}
&W_t+gW(D_1D_2W)_x+h(D_1W)(D_2W)_x+i(D_2W)(D_1W)_x\\&+(h-i-g-1)(D_1D_2W)W_x+j(D_1D_2P)^{\gamma}P_x+k(D_1D_2P)^{\gamma-1}(D_2P)(D_1P)_x\\&+l(D_1D_2P)^{\gamma-1}(D_1P)(D_2P)_x+m(D_1D_2P)^{\gamma-1}P(D_1D_2P)_x\\&+\left((-1)^{\gamma+1}A+j+k+m-l\right)(D_1D_2P)^{\gamma-2}(D_1P)(D_2P)(D_1D_2P)_x=0
\end{split}
\label{b6}
\end{equation}
where $a$, $b$, ... , $m$ are thirteen arbitrary bosonic parameters. In this paper, we focus on the simplest case where all of the parameters vanish:
\begin{equation}
\begin{split}
\Delta_1\equiv &P_t-P_x(D_1D_2W)-(D_1D_2P)W_x=0\\
& \\
\Delta_2\equiv &W_t-(D_1D_2W)W_x+(-1)^{\gamma+1}A(D_1D_2P)^{\gamma-2}(D_1P)(D_2P)(D_1D_2P)_x=0
\end{split}
\label{c1}
\end{equation}
We will subsequently refer to system (\ref{c1}) as the supersymmetric polytropic gas dynamics equations.

A real Grassmann algebra $\mathfrak{G}$ is generated by a finite or infinite number of generators $(\xi_1,\xi_2,\ldots)$. It is a graded vector space $\mathfrak{G}=\mathfrak{G}_0+\mathfrak{G}_1$, where each variable $Y$ in $\mathfrak{G}$ that we use is either even or odd. In general, for a Grassmann variable $Y$, we can define a ``parity'' $\tilde{Y}$ which is $0$ if $Y$ is even and $1$ if $Y$ is odd. In order to derive equation (\ref{c1}), we have used the following Leibniz rule:
\begin{equation}
\label{gLr}
\partial_{\theta_i} (fg) = (\partial_{\theta_i} f)g+(-1)^{\tilde f} (\partial_{\theta_i} g).
\end{equation}
The partial derivatives with respect to odd coordinate satisfy the usual operational rules, namely $\partial_{\theta_i} \theta_j=\delta^i_j$. The operators $\partial_{\theta_i}$, $Q_1$, $Q_2$, $D_1$ and $D_2$ change the ``parity'' of the function acted on in the sense that it converts a bosonic function to a fermionic function and vice-versa. For example, $\partial_{\theta_i}W$ is an odd superfield while $\partial_{\theta_1}\partial_{\theta_2}W$ is an even superfield, and so on. For further details, see the book by Cornwell \cite{Cornwell} and the reference by DeWitt \cite{DeWitt}.

\section{Symmetries of the supersymmetric polytropic gas dynamics equation}

In order to determine the Lie superalgebra of infinitesimal symmetries of the supersymmetric equations (\ref{c1}), we make use of the apparatus of vector field prolongations as described in the book by P. J. Olver \cite{Olver}. Specifically, we make use of a generalization of this method to Grassmann variables in analogy with the case of the supersymmetric sine-Gordon equation \cite{SSG}.

A symmetry supergroup $G$ of the system (\ref{c1}) is a (local) supergroup of transformations acting on the cartesian product of supermanifolds 
$X\times U$,
where $X$ represents the independent variables $(x,t,\theta_1,\theta_2)$ and $U$ the dependent variables (superfields) $(W,P)$.
The action of $G$ on the functions $W(x,t,\theta_1,\theta_2)$ and $P(x,t,\theta_1,\theta_2)$ maps solutions of (\ref{c1}) to solutions of (\ref{c1}). Assuming that $G$ is a Lie supergroup as described in \cite{ARG,ARG2,Roelofs2} one can associate to it its Lie superalgebra of even left--invariant vector fields $\mathfrak{g}$, whose elements are the infinitesimal symmetries of the system (\ref{c1}). In particular, a local one-parameter subgroup of $G$ consists of a family of transformations
\begin{equation}
g_{\varepsilon}:\quad \tilde{x}^i = X^i(x,u,\varepsilon),\quad \tilde{u}^{\alpha} = U^{\alpha}(x,u,\varepsilon)
\end{equation}
where $x=(x^1,x^2,x^3,x^4)=(x,t,\theta_1,\theta_2)$ are the independent variables and $u=(u^1,u^2)=(W,P)$ the dependent variables. Here, $\varepsilon$ is a real parameter, whose range may be restricted depending on the values of $x,t,\theta_1,\theta_2,W,P$. Such a local subgroup is generated by a vector field of the form
\begin{equation}
\mathbf{v} = \xi^i(x,u){\partial\over \partial x^i} + \Phi(x,u)^{\alpha}{\partial\over \partial u^{\alpha}},
\label{thisfield1}
\end{equation}
where the coefficients are given by
\begin{equation}
\xi^i(x,u) = {d\over d\varepsilon}X^i|_{\varepsilon = 0},\quad \Phi^{\alpha}(x,u) = {d\over d\varepsilon}U^{\alpha}|_{\varepsilon = 0}.
\end{equation}
The Lie superalgebra $\mathfrak{g}$, consisting of all vector fields of the form (\ref{thisfield1}), then generates the Lie supergroup $G$. The advantage of working with the Lie superalgebra $\mathfrak{g}$ instead of directly with the Lie supergroup $G$ is that {the equations defining} the infinitesimal symmetries are linear.

In order to determine the infinitesimal symmetries of a system of partial differential equations, it is useful to make use of the concept of the prolongation of a (super)group action. The idea is that the transformation  of the coordinates $x^i\rightarrow\tilde{x}^i$, $u^{\alpha}\rightarrow\tilde{u}^{\alpha}$ induces a transformation of the derivatives
\begin{equation}
{\partial u^{\alpha}\over \partial x^i}\longrightarrow {\partial \tilde{u}^{\alpha}\over \partial \tilde{x}^i}.
\end{equation}
In order to make use of this concept, we define the multi-index $J = (j_1,\ldots,j_p)$, where $j_i = 0,1,\ldots$ and
$|J| = j_1 + \ldots + j_p$. 
The space of coordinates on $X\times U$ is extended to the jet bundle
%\begin{equation}
${\mathcal J}_k = \{(x^i,u^{\alpha},u^{\alpha}_J)| \, |J|\leq k\}$,
%\end{equation}
which includes the coordinates and all derivatives of the dependent variables of order less than or equal to $k$.  In our setting the jet bundle is a supermanifold as well, since $X\times U$ was. On the jet bundle, we define total derivatives
\begin{equation} {\mathcal D}_i = {\partial\over \partial x^i} +
  \sum\limits_{\alpha,J}u^{\alpha}_{J_i}{\partial\over \partial
    u^{\alpha}_J}\mbox{,}\end{equation}
where $J_i = (j_1,\ldots,j_{i-1},j_i+1,j_{i+1},\ldots,j_n)$.
More generally, for $J = (j_1,j_2,\ldots,j_n)$, we define the composite derivative
\begin{equation}{\mathcal D}_J = \underbrace{{\mathcal D}_1{\mathcal D}_1\cdots {\mathcal D}_1}_{j_1} \cdots\ \underbrace{{\mathcal D}_n{\mathcal D}_n\cdots {\mathcal D}_n}_{j_n}. \end{equation} 

The prolongation of a Lie (super)group action to the jet bundle ${\mathcal J}_k$ in turn induces a prolongation of the generating infinitesimal vector field in the Lie (super)algebra. For the vector field $\mathbf{v}$ given by (\ref{thisfield1}), the $k^{th}$ order prolongation of the vector field $\mathbf{v}$ is
\begin{equation}pr^{(k)}(\mathbf{v}) = \mathbf{v} + \sum\limits_{\alpha,|J|\neq 0}\phi^{\alpha}_J(x,u^{(k)}){\partial\over \partial u^{\alpha}_J}\mbox{,} \end{equation}
where  $\phi^{\alpha}_J(x,u^{(k)})$ are given by the formula
\begin{equation}\label{xprolong}\phi^{\alpha}_J = {\mathcal D}_J\left(\phi^{\alpha} - \xi^i{\partial u^{\alpha}\over \partial x^i}\right) + \xi^i u^{\alpha}_{J_i},
\end{equation}
or, equivalently, by the recursive formula
\begin{equation}\label{prbos}
 \phi^{\alpha}_{J_j}={\mathcal D}_{j} \phi^\alpha_J - \sum_{i} ({\mathcal D}_{j} \xi^i) u^{\alpha}_{J_i}.
\end{equation}
The symmetry criterion (Theorem 2.31 in Olver \cite{Olver}) assumes that $G$ is a connected Lie group of transformations acting locally on $X\times
  U$ through the transformations
\begin{displaymath}\tilde{x}_i = X^i(x,u,g), \qquad \tilde{u}^{\alpha} = U^{\alpha}(x,u,g)\mbox{,}\end{displaymath}
where $g\in G$ and $\Delta_{\nu}(x,u^{(n)})$ is a non--degenerate system of partial differential equations (meaning that the system is locally solvable and is of maximal rank at every point $(x_0,u_0^{(n)})\in X\times U^{(n)}$). Then $G$ is a symmetry group of $\Delta = 0$ if and only if
\begin{equation}\left[pr^{(k)}(\mathbf{v})\right](\Delta) = 0 \qquad  \mbox{whenever} \qquad  \Delta = 0\mbox{,}\end{equation}
for each infinitesimal generator $\mathbf{v}$ of $G$.
Using the results of \cite{ARG,ARG2}, one finds that the same criterion can be used also in the case of the Lie supergroup $G$ and its Lie superalgebra of even left--invariant vector fields.

For the purpose of determining the Lie superalgebra of symmetries of the system (\ref{c1}), let us write a vector field of the form
\begin{equation}
\begin{split}
\mathbf{v}=&\xi(x,t,\theta_1,\theta_2,W,P)\partial_x+\tau(x,t,\theta_1,\theta_2,W,P)\partial_t+\rho_1(x,t,\theta_1,\theta_2,W,P)\partial_{\theta_1}\\ &+\rho_2(x,t,\theta_1,\theta_2,W,P)\partial_{\theta_2}+\Lambda(x,t,\theta_1,\theta_2,W,P)\partial_W+\Omega(x,t,\theta_1,\theta_2,W,P)\partial_P,
\end{split}
\label{symmie3}
\end{equation}
where $\xi$, $\tau$, $\Lambda$ and $\Omega$ are bosonic functions, while $\rho_1$ and $\rho_2$ are fermionic. In general, we use the convention that the fermionic coefficients in a vector field expansion (in this case $\rho_1$ and $\rho_2$) precede the fermionic derivatives which they are multiplied by ( in this case $\partial_{\theta_1}$ and $\partial_{\theta_2}$ respectively).

According to the symmetry criterion, the vector field (\ref{symmie3}) is an infinitesimal generator of the symmetry group of the system of differential equations (\ref{c1}) if and only if
\begin{equation}
\mbox{pr}^{(2)}(\mathbf{v})\left[\Delta_k(x,t,\theta_1,\theta_2,W,P,W^{(2)},P^{(2)})\right]=0,\quad k=1,2
\label{c1E}
\end{equation}
whenever $\Delta_l(x,t,\theta_1,\theta_2,W,P,W^{(2)},P^{(2)})=0$, $l=1,2$.
We use the following expressions for the total derivatives ${\mathcal D}_x$, ${\mathcal D}_t$, ${\mathcal D}_{\theta_1}$ and ${\mathcal D}_{\theta_2}$:
\begin{equation}
\begin{split}
{\mathcal D}_x=&\partial_x+W_x\partial_{W}+P_x\partial_{P}+W_{xx}\partial_{W_x}+P_{xx}\partial_{P_x}+W_{xt}\partial_{W_t}+P_{xt}\partial_{P_t}+W_{x\theta_1}\partial_{W_{\theta_1}}\\
&+P_{x\theta_1}\partial_{P_{\theta_1}}+W_{x\theta_2}\partial_{W_{\theta_2}}+P_{x\theta_2}\partial_{P_{\theta_2}}+W_{xxx}\partial_{W_{xx}}+P_{xxx}\partial_{P_{xx}}+W_{xxt}\partial_{W_{xt}}\\ &+P_{xxt}\partial_{P_{xt}}+W_{xx\theta_1}\partial_{W_{x\theta_1}}+P_{xx\theta_1}\partial_{P_{x\theta_1}}+W_{xx\theta_2}\partial_{W_{x\theta_2}}+P_{xx\theta_2}\partial_{P_{x\theta_2}}+W_{xtt}\partial_{W_{tt}}\\ &+P_{xtt}\partial_{P_{tt}}+W_{xt\theta_1}\partial_{W_{t\theta_1}}+P_{xt\theta_1}\partial_{P_{t\theta_1}}+W_{xt\theta_2}\partial_{W_{t\theta_2}}+P_{xt\theta_2}\partial_{P_{t\theta_2}}+W_{x\theta_1\theta_2}\partial_{W_{\theta_1\theta_2}}\\ &+P_{x\theta_1\theta_2}\partial_{P_{\theta_1\theta_2}},
\end{split}
\label{symmie4}
\end{equation}
\begin{equation}
\begin{split}
{\mathcal D}_t=&\partial_t+W_t\partial_{W}+P_t\partial_{P}+W_{xt}\partial_{W_x}+P_{xt}\partial_{P_x}+W_{tt}\partial_{W_t}+P_{tt}\partial_{P_t}+W_{t\theta_1}\partial_{W_{\theta_1}}\\ &+P_{t\theta_1}\partial_{P_{\theta_1}}+W_{t\theta_2}\partial_{W_{\theta_2}}+P_{t\theta_2}\partial_{P_{\theta_2}}+W_{xxt}\partial_{W_{xx}}+P_{xxt}\partial_{P_{xx}}+W_{xtt}\partial_{W_{xt}}\\ &+P_{xtt}\partial_{P_{xt}}+W_{xt\theta_1}\partial_{W_{x\theta_1}}+P_{xt\theta_1}\partial_{P_{x\theta_1}}+W_{xt\theta_2}\partial_{W_{x\theta_2}}+P_{xt\theta_2}\partial_{P_{x\theta_2}}+W_{ttt}\partial_{W_{tt}}\\ &+P_{ttt}\partial_{P_{tt}}+W_{tt\theta_1}\partial_{W_{t\theta_1}}+P_{tt\theta_1}\partial_{P_{t\theta_1}}+W_{tt\theta_2}\partial_{W_{t\theta_2}}+P_{tt\theta_2}\partial_{P_{t\theta_2}}+W_{t\theta_1\theta_2}\partial_{W_{\theta_1\theta_2}}\\ &+P_{t\theta_1\theta_2}\partial_{P_{\theta_1\theta_2}},
\end{split}
\label{symmie4.5}
\end{equation}
\begin{equation}
\begin{split}
{\mathcal D}_{\theta_1}=&\partial_{\theta_1}+W_{\theta_1}\partial_{W}+P_{\theta_1}\partial_{P}+W_{x\theta_1}\partial_{W_x}+P_{x\theta_1}\partial_{P_x}+W_{t\theta_1}\partial_{W_t}+P_{t\theta_1}\partial_{P_t}+W_{\theta_2\theta_1}\partial_{W_{\theta_2}}\\ &+P_{\theta_2\theta_1}\partial_{P_{\theta_2}}+W_{xx\theta_1}\partial_{W_{xx}}+P_{xx\theta_1}\partial_{P_{xx}}+W_{xt\theta_1}\partial_{W_{xt}}+P_{xt\theta_1}\partial_{P_{xt}}+W_{x\theta_2\theta_1}\partial_{W_{x\theta_2}}\\ &+P_{x\theta_2\theta_1}\partial_{P_{x\theta_2}}+W_{tt\theta_1}\partial_{W_{tt}}+P_{tt\theta_1}\partial_{P_{tt}}+W_{t\theta_2\theta_1}\partial_{W_{t\theta_2}}+P_{t\theta_2\theta_1}\partial_{P_{t\theta_2}},
\end{split}
\label{symmie5}
\end{equation}
and
\begin{equation}
\begin{split}
{\mathcal D}_{\theta_2}=&\partial_{\theta_2}+W_{\theta_2}\partial_{W}+P_{\theta_2}\partial_{P}+W_{x\theta_2}\partial_{W_x}+P_{x\theta_2}\partial_{P_x}+W_{t\theta_2}\partial_{W_t}+P_{t\theta_2}\partial_{P_t}+W_{\theta_1\theta_2}\partial_{W_{\theta_1}}\\ &+P_{\theta_1\theta_2}\partial_{P_{\theta_1}}+W_{xx\theta_2}\partial_{W_{xx}}+P_{xx\theta_2}\partial_{P_{xx}}+W_{xt\theta_2}\partial_{W_{xt}}+P_{xt\theta_2}\partial_{P_{xt}}+W_{x\theta_1\theta_2}\partial_{W_{x\theta_1}}\\ &+P_{x\theta_1\theta_2}\partial_{P_{x\theta_1}}+W_{tt\theta_2}\partial_{W_{tt}}+P_{tt\theta_2}\partial_{P_{tt}}+W_{t\theta_1\theta_2}\partial_{W_{t\theta_1}}+P_{t\theta_1\theta_2}\partial_{P_{t\theta_1}},
\end{split}
\label{symmie5.5}
\end{equation}

We note that the chain rule for a Grassmann-valued composite function $f(g(x))$ is \cite{Berezin,DeWitt}
\begin{equation}
{\partial f\over \partial x}={\partial g\over \partial x}\cdot{\partial f\over \partial g}.
\label{chainrule}
\end{equation}
The interchangeability of mixed derivatives (with proper respect to the ordering of odd variables) is of course assumed throughout. The second prolongation of the vector field (\ref{symmie3}) is given by
\begin{equation}
\begin{split}
\mbox{pr}^{(2)}\mathbf{v}=&\xi\partial_x+\tau\partial_t+\rho_1\partial_{\theta_1}+\rho_2\partial_{\theta_2}+\Lambda\partial_{W}+\Omega\partial_{P}+\Lambda^x\partial_{W_x}+\Lambda^t\partial_{W_t}+\Lambda^{\theta_1}\partial_{W_{\theta_1}}\\ &+\Lambda^{\theta_2}\partial_{W_{\theta_2}}+\Lambda^{xx}\partial_{W_{xx}}+\Lambda^{xt}\partial_{W_{xt}}+\Lambda^{x\theta_1}\partial_{W_{x\theta_1}}+\Lambda^{x\theta_2}\partial_{W_{x\theta_2}}+\Lambda^{tt}\partial_{W_{tt}}\\ &+\Lambda^{t\theta_1}\partial_{W_{t\theta_1}}+\Lambda^{t\theta_2}\partial_{W_{t\theta_2}}+\Lambda^{\theta_1\theta_2}\partial_{W_{\theta_1\theta_2}}
+\Omega^x\partial_{P_x}+\Omega^t\partial_{P_t}+\Omega^{\theta_1}\partial_{P_{\theta_1}}\\ &+\Omega^{\theta_2}\partial_{P_{\theta_2}}+\Omega^{xx}\partial_{P_{xx}}+\Omega^{xt}\partial_{P_{xt}}+\Omega^{x\theta_1}\partial_{P_{x\theta_1}}+\Omega^{x\theta_2}\partial_{P_{x\theta_2}}+\Omega^{tt}\partial_{P_{tt}}\\ &+\Omega^{t\theta_1}\partial_{P_{t\theta_1}}+\Omega^{t\theta_2}\partial_{P_{t\theta_2}}+\Omega^{\theta_1\theta_2}\partial_{P_{\theta_1\theta_2}}.
\end{split}
\label{symmie6}
\end{equation}
Here, we use upper indices in coefficients $\Lambda^x$, $\Omega^{x\theta_1}$ etc. in order to distinguish them from partial derivatives, e.g. $\Omega_{x\theta_1} = \partial_{\theta_1}\partial_{x} \Omega$.
Applying the second prolongation (\ref{symmie6}) to the equation (\ref{c1}), we obtain the following conditions for the coefficients of the second prolongation (\ref{symmie6})
\begin{equation}
\begin{split}
& \rho_1\partial_{\theta_1}\left[P_t-P_x(D_1D_2W)-(D_1D_2P)W_x\right]\\ &+\rho_2\partial_{\theta_2}\left[P_t-P_x(D_1D_2W)-(D_1D_2P)W_x\right]\\ & +\Omega^x\partial_{P_x}\left[P_t-P_x(D_1D_2W)-(D_1D_2P)W_x\right]\\ &+\Omega^t\partial_{P_t}\left[P_t-P_x(D_1D_2W)-(D_1D_2P)W_x\right]\\ & +\Lambda^{xt}\partial_{W_{xt}}\left[P_t-P_x(D_1D_2W)-(D_1D_2P)W_x\right]\\ & +\Lambda^{t\theta_1}\partial_{W_{t\theta_1}}\left[P_t-P_x(D_1D_2W)-(D_1D_2P)W_x\right]\\ & +\Lambda^{x\theta_2}\partial_{W_{x\theta_2}}\left[P_t-P_x(D_1D_2W)-(D_1D_2P)W_x\right]\\ & +\Lambda^{\theta_1\theta_2}\partial_{W_{\theta_1\theta_2}}\left[P_t-P_x(D_1D_2W)-(D_1D_2P)W_x\right]\\ & +\Omega^{xt}\partial_{P_{xt}}\left[P_t-P_x(D_1D_2W)-(D_1D_2P)W_x\right]\\ & +\Omega^{t\theta_1}\partial_{P_{t\theta_1}}\left[P_t-P_x(D_1D_2W)-(D_1D_2P)W_x\right]\\ & +\Omega^{x\theta_2}\partial_{P_{x\theta_2}}\left[P_t-P_x(D_1D_2W)-(D_1D_2P)W_x\right]\\ & +\Omega^{\theta_1\theta_2}\partial_{P_{\theta_1\theta_2}}\left[P_t-P_x(D_1D_2W)-(D_1D_2P)W_x\right]=0,
\end{split}
\label{cccccc1}
\end{equation}
and
\begin{equation}
\begin{split}
& \rho_1\partial_{\theta_1}\left[W_t-(D_1D_2W)W_x+(-1)^{\gamma+1}A(D_1D_2P)^{\gamma-2}(D_1P)(D_2P)(D_1D_2P)_x\right]\\ & +\rho_2\partial_{\theta_2}\left[W_t-(D_1D_2W)W_x+(-1)^{\gamma+1}A(D_1D_2P)^{\gamma-2}(D_1P)(D_2P)(D_1D_2P)_x\right]\\ & +\Omega^x\partial_{P_x}\left[W_t-(D_1D_2W)W_x+(-1)^{\gamma+1}A(D_1D_2P)^{\gamma-2}(D_1P)(D_2P)(D_1D_2P)_x\right]\\ & +\Omega^t\partial_{P_t}\left[W_t-(D_1D_2W)W_x+(-1)^{\gamma+1}A(D_1D_2P)^{\gamma-2}(D_1P)(D_2P)(D_1D_2P)_x\right]\\ & +\Lambda^{xt}\partial_{W_{xt}}\left[W_t-(D_1D_2W)W_x+(-1)^{\gamma+1}A(D_1D_2P)^{\gamma-2}(D_1P)(D_2P)(D_1D_2P)_x\right]\\ & +\Lambda^{t\theta_1}\partial_{W_{t\theta_1}}\left[W_t-(D_1D_2W)W_x+(-1)^{\gamma+1}A(D_1D_2P)^{\gamma-2}(D_1P)(D_2P)(D_1D_2P)_x\right]\\ & +\Lambda^{x\theta_2}\partial_{W_{x\theta_2}}\left[W_t-(D_1D_2W)W_x+(-1)^{\gamma+1}A(D_1D_2P)^{\gamma-2}(D_1P)(D_2P)(D_1D_2P)_x\right]\\ & +\Lambda^{\theta_1\theta_2}\partial_{W_{\theta_1\theta_2}}\left[W_t-(D_1D_2W)W_x+(-1)^{\gamma+1}A(D_1D_2P)^{\gamma-2}(D_1P)(D_2P)(D_1D_2P)_x\right]\\ & +\Omega^{xt}\partial_{P_{xt}}\left[W_t-(D_1D_2W)W_x+(-1)^{\gamma+1}A(D_1D_2P)^{\gamma-2}(D_1P)(D_2P)(D_1D_2P)_x\right]\\ & +\Omega^{t\theta_1}\partial_{P_{t\theta_1}}\left[W_t-(D_1D_2W)W_x+(-1)^{\gamma+1}A(D_1D_2P)^{\gamma-2}(D_1P)(D_2P)(D_1D_2P)_x\right]\\ & +\Omega^{x\theta_2}\partial_{P_{x\theta_2}}\left[W_t-(D_1D_2W)W_x+(-1)^{\gamma+1}A(D_1D_2P)^{\gamma-2}(D_1P)(D_2P)(D_1D_2P)_x\right]\\ & +\Omega^{\theta_1\theta_2}\partial_{P_{\theta_1\theta_2}}\left[W_t-(D_1D_2W)W_x+(-1)^{\gamma+1}A(D_1D_2P)^{\gamma-2}(D_1P)(D_2P)(D_1D_2P)_x\right]\\ & =0.
\end{split}
\label{cccccc2}
\end{equation}

Note that proper respect to the ordering of fermionic terms is essential, e.g. $\Lambda^{t\theta_1}$ is odd. We see that we only need to calculate the coefficients $\Omega^x$, $\Omega^t$, $\Lambda^{xt}$, $\Lambda^{t\theta_1}$, $\Lambda^{x\theta_2}$, $\Lambda^{\theta_1\theta_2}$, $\Omega^{xt}$, $\Omega^{t\theta_1}$, $\Omega^{x\theta_2}$, $\Omega^{\theta_1\theta_2}$ in equation (\ref{symmie6}). They are found from the superspace version of the formulae for the 1st and 2nd prolongation (\ref{xprolong})
\begin{equation}
 \Lambda^{A}={\mathcal D}_{A} \Lambda - \sum_{B} {\mathcal D}_{A} \zeta^B \Phi_B, \qquad 
\Lambda^{AB}={\mathcal D}_{B} \Lambda^A - \sum_{C} {\mathcal D}_{B} \zeta^C \Phi_{AC},
\label{symmie7A}
\end{equation}
where
\begin{equation}
 A,B,C\in\{x,t,\theta_1,\theta_2\}, \qquad \zeta^A=(\xi,\tau,\rho,\sigma).
\end{equation}
The derivation of these formulae is performed in the same way as in the bosonic case, working formally with infinitesimal transformations and keeping track of ordering properties. The signs in the expressions below vary according to the considered equation (\ref{c1}).
Explicitly, the coefficients of the prolongation (\ref{symmie6}) are given as follows:
\begin{equation}
\begin{split}
\Omega^x=&\Omega_x+\Omega_{W}W_x+\Omega_{P}P_x-\xi_x P_x-\xi_{W}W_xP_x-\xi_{P}(P_x)^2-\tau_x P_t-\tau_WW_xP_t\\ &-\tau_{P}P_x P_t-\rho_{1,x}P_{\theta_1}-\rho_{1,W}W_xP_{\theta_1}-\rho_{1,P}P_xP_{\theta_1}-\rho_{2,x}P_{\theta_2}-\rho_{2,W}W_xP_{\theta_2}\\ &-\rho_{2,P}P_xP_{\theta_2},
\end{split}
\label{littlecoeff1}
\end{equation}

\begin{equation}
\begin{split}
\Omega^t=&\Omega_t+\Omega_{W}W_t+\Omega_{P}P_t-\xi_tP_x-\xi_{W}W_tP_x-\xi_{P}P_xP_t-\tau_tP_t-\tau_{W}W_tP_t\\ &-\tau_{P}(P_t)^2-\rho_{1,t}P_{\theta_1}-\rho_{1,W}W_tP_{\theta_1}-\rho_{1,P}P_tP_{\theta_1}-\rho_{2,t}P_{\theta_2}-\rho_{2,W}W_tP_{\theta_2}\\ &-\rho_{2,P}P_tP_{\theta_2},
\end{split}
\label{littlecoeff2}
\end{equation}

\begin{equation}
\begin{split}
\Lambda^{xt}=&\Lambda_{xt}+\Lambda_{xW}W_t+\Lambda_{xP}P_t+\Lambda_{tW}W_x+\Lambda_{WW}W_xW_t+\Lambda_{WP}W_xP_t+\Lambda_{W}W_{xt}\\ &+\Lambda_{tP}P_x+\Lambda_{WP}P_xW_t+\Lambda_{PP}P_xP_t+\Lambda_{P}P_{xt}-\xi_{xt}W_x-\xi_{xW}W_xW_t-\xi_{xP}W_xP_t\\ &-\xi_xW_{xt}-\xi_{tW}(W_x)^2-\xi_{WW}(W_x)^2W_t-\xi_{WP}(W_x)^2P_t-2\xi_{W}W_xW_{xt}\\ &-\xi_{tP}W_xP_x-\xi_{WP}W_xW_tP_x-\xi_{PP}W_xP_xP_t-\xi_{P}P_xW_{xt}-\xi_{P}W_xP_{xt}-\xi_tW_{xx}\\ &-\xi_{W}W_tW_{xx}-\xi_{P}P_tW_{xx}-\tau_{xt}W_t-\tau_{xW}(W_t)^2-\tau_{xP}W_tP_t-\tau_xW_{tt}-\tau_{tW}W_xW_t\\ &-\tau_{WW}(W_t)^2W_x-\tau_{WP}W_xW_tP_t-2\tau_{W}W_tW_{xt}-\tau_{W}W_xW_{tt}-\tau_{tP}W_tP_x\\ &-\tau_{WP}(W_t)^2P_x-\tau_{PP}W_tP_xP_t-\tau_{P}P_xW_{tt}-\tau_{P}W_tP_{xt}-\tau_tW_{xt}-\tau_{P}P_tW_{xt}\\ 
&-\rho_{1,xt}W_{\theta_1}-\rho_{1,xW}W_tW_{\theta_1}-\rho_{1,xP}P_tW_{\theta_1}-\rho_{1,tW}W_xW_{\theta_1}-\rho_{1,x}W_{t\theta_1}-\rho_{1,t}W_{x\theta_1}\\ &-\rho_{1,WW}W_xW_tW_{\theta_1}-\rho_{1,WP}W_xW_{\theta_1}P_t-\rho_{1,W}W_{xt}W_{\theta_1}-\rho_{1,W}W_{t\theta_1}W_x\\ &-\rho_{1,W}W_{x\theta_1}W_t-\rho_{1,tP}P_xW_{\theta_1}-\rho_{1,WP}P_xW_tW_{\theta_1}-\rho_{1,PP}P_xP_tW_{\theta_1}-\rho_{1,P}P_{xt}W_{\theta_1}\\ &-\rho_{1,P}P_xW_{t\theta_1}-\rho_{1,P}P_tW_{x\theta_1}-\rho_{2,xt}W_{\theta_2}-\rho_{2,xW}W_tW_{\theta_2}-\rho_{2,xP}P_tW_{\theta_2}\\ &-\rho_{2,tW}W_xW_{\theta_2}-\rho_{2,x}W_{t\theta_2}-\rho_{2,t}W_{x\theta_2}-\rho_{2,WW}W_xW_tW_{\theta_2}-\rho_{2,WP}P_tW_xW_{\theta_2}\\ &-\rho_{2,W}W_{xt}W_{\theta_2}-\rho_{2,W}W_{t\theta_2}W_x-\rho_{2,W}W_{x\theta_2}W_t-\rho_{2,tP}P_xW_{\theta_2}-\rho_{2,WP}P_xW_tW_{\theta_2}\\ &-\rho_{2,PP}P_xP_tW_{\theta_2}-\rho_{2,P}P_{xt}W_{\theta_2}-\rho_{2,P}W_{t\theta_2}P_x-\rho_{2,P}W_{x\theta_2}P_t,\end{split}
\label{bigcoeff1}
\end{equation}

\begin{equation}
\begin{split}
\Lambda^{t\theta_1}=&\Lambda_{t\theta_1}+\Lambda_{tW}W_{\theta_1}+\Lambda_{tP}P_{\theta_1}+\Lambda_{\theta_1W}W_t+\Lambda_{WW}W_tW_{\theta_1}+\Lambda_{WP}W_tP_{\theta_1}+\Lambda_{W}W_{t\theta_1}\\ &+\Lambda_{\theta_1P}P_t+\Lambda_{WP}P_tW_{\theta_1}+\Lambda_{PP}P_tP_{\theta_1}+\Lambda_{P}P_{t\theta_1}
-\xi_{t\theta_1}W_x-\xi_{tW}W_xW_{\theta_1}\\ &-\xi_{tP}W_xP_{\theta_1}-\xi_tW_{x\theta_1}-\xi_{\theta_1W}W_xW_t-\xi_{WW}W_xW_tW_{\theta_1}-\xi_{WP}W_xW_tP_{\theta_1}\\ &-\xi_{W}W_tW_{x\theta_1}-\xi_{W}W_xW_{t\theta_1}-\xi_{\theta_1P}W_xP_t-\xi_{WP}P_tW_xW_{\theta_1}-\xi_{PP}W_xP_tP_{\theta_1}\\ &-\xi_{P}P_tW_{x\theta_1}-\xi_{P}W_xP_{t\theta_1}-\xi_{\theta_1}W_{xt}-\xi_{W}W_{xt}W_{\theta_1}-\xi_{P}W_{xt}P_{\theta_1}-\tau_{t\theta_1}W_t\\ &-\tau_{tW}W_tW_{\theta_1}-\tau_{tP}W_tP_{\theta_1}-\tau_tW_{t\theta_1}-\tau_{\theta_1W}(W_t)^2-\tau_{WW}(W_t)^2W_{\theta_1}\\ &-\tau_{WP}(W_t)^2P_{\theta_1}-2\tau_{W}W_tW_{t\theta_1}-\tau_{\theta_1P}W_tP_t-\tau_{WP}P_tW_tW_{\theta_1}-\tau_{PP}W_tP_tP_{\theta_1}\\ &-\tau_{P}P_tW_{t\theta_1}-\tau_{P}W_tP_{t\theta_1}-\tau_{\theta_1}W_{tt}-\tau_{W}W_{tt}W_{\theta_1}-\tau_{P}W_{tt}P_{\theta_1}-\rho_{1,t\theta_1}W_{\theta_1}\\ &+\rho_{1,tP}P_{\theta_1}W_{\theta_1}-\rho_{1,\theta_1W}W_tW_{\theta_1}+\rho_{1,WP}P_{\theta_1}W_tW_{\theta_1}+\rho_{1,W}W_{t\theta_1}W_{\theta_1}\\ &-\rho_{1,\theta_1P}P_tW_{\theta_1}+\rho_{1,PP}P_tP_{\theta_1}W_{\theta_1}+\rho_{1,P}P_{t\theta_1}W_{\theta_1}-\rho_{1,\theta_1}W_{t\theta_1}+\rho_{1,W}W_{\theta_1}W_{t\theta_1}\\ &+\rho_{1,P}P_{\theta_1}W_{t\theta_1}-\rho_{2,t\theta_1}W_{\theta_2}+\rho_{2,tW}W_{\theta_1}W_{\theta_2}+\rho_{2,tP}P_{\theta_1}W_{\theta_2}-\rho_{2,t}W_{\theta_1\theta_2}\\ &-\rho_{2,\theta_1W}W_tW_{\theta_2}+\rho_{2,WW}W_tW_{\theta_1}W_{\theta_2}+\rho_{2,WP}P_{\theta_1}W_tW_{\theta_2}+\rho_{2,W}W_{t\theta_1}W_{\theta_2}\\ &-\rho_{2,W}W_tW_{\theta_1\theta_2}-\rho_{2,\theta_1P}P_tW_{\theta_2}+\rho_{2,WP}P_tW_{\theta_1}W_{\theta_2}+\rho_{2,PP}P_tP_{\theta_1}W_{\theta_2}\\ &+\rho_{2,P}P_{t\theta_1}W_{\theta_2}-\rho_{2,P}P_tW_{\theta_1\theta_2}-\rho_{2,\theta_1}W_{t\theta_2}+\rho_{2,W}W_{\theta_1}W_{t\theta_2}+\rho_{2,P}P_{\theta_1}W_{t\theta_2},
\end{split}
\label{bigcoeff2}
\end{equation}

\begin{equation}
\begin{split}
\Lambda^{x\theta_2}=&\Lambda_{x\theta_2}+\Lambda_{xW}W_{\theta_2}+\Lambda_{xP}P_{\theta_2}+\Lambda_{\theta_2W}W_x+\Lambda_{WW}W_xW_{\theta_2}+\Lambda_{WP}W_xP_{\theta_2}+\Lambda_{W}W_{x\theta_2}\\ &+\Lambda_{\theta_2P}P_x+\Lambda_{WP}P_xW_{\theta_2}+\Lambda_{PP}P_xP_{\theta_2}+\Lambda_{P}P_{x\theta_2}-\xi_{x\theta_2}W_x-\xi_{xW}W_xW_{\theta_2}-\xi_{xP}W_xP_{\theta_2}\\ &-\xi_xW_{x\theta_2}-\xi_{\theta_2W}(W_x)^2-\xi_{WW}(W_x)^2W_{\theta_2}-\xi_{WP}(W_x)^2P_{\theta_2}-2\xi_{W}W_xW_{x\theta_2}\\ &-\xi_{\theta_2P}W_xP_x-\xi_{WP}P_xW_xW_{\theta_2}-\xi_{PP}W_xP_xP_{\theta_2}-\xi_{P}P_xW_{x\theta_2}-\xi_{P}W_xP_{x\theta_2}\\ &-\xi_{\theta_2}W_{xx}-\xi_{W}W_{xx}W_{\theta_2}-\xi_{P}W_{xx}P_{\theta_2}-\tau_{x\theta_2}W_t-\tau_{xW}W_tW_{\theta_2}-\tau_{xP}W_tP_{\theta_2}\\ &-\tau_xW_{t\theta_2}-\tau_{\theta_2W}W_xW_t-\tau_{WW}W_xW_tW_{\theta_2}-\tau_{WP}W_xW_tP_{\theta_2}-\tau_{W}W_xW_{t\theta_2}\\ &-\tau_{W}W_tW_{x\theta_2}-\tau_{\theta_2P}W_tP_x-\tau_{WP}P_xW_tW_{\theta_2}-\tau_{PP}W_tP_xP_{\theta_2}-\tau_{P}P_xW_{t\theta_2}\\ &-\tau_{P}W_tP_{x\theta_2}-\tau_{\theta_2}W_{xt}-\tau_{W}W_{xt}W_{\theta_2}-\tau_{P}W_{xt}P_{\theta_2}-\rho_{1,x\theta_2}W_{\theta_1}+\rho_{1,xW}W_{\theta_2}W_{\theta_1}\\ &+\rho_{1,xP}P_{\theta_2}W_{\theta_1}+\rho_{1,x}W_{\theta_1\theta_2}-\rho_{1,\theta_2W}W_xW_{\theta_1}+\rho_{1,WW}W_xW_{\theta_2}W_{\theta_1}\\ &+\rho_{1,WP}P_{\theta_2}W_xW_{\theta_1}+\rho_{1,W}W_{x\theta_2}W_{\theta_1}+\rho_{1,W}W_xW_{\theta_1\theta_2}-\rho_{1,\theta_2P}P_xW_{\theta_1}\\ &+\rho_{1,WP}P_xW_{\theta_2}W_{\theta_1}+\rho_{1,PP}P_xP_{\theta_2}W_{\theta_1}+\rho_{1,P}P_{x\theta_2}W_{\theta_1}+\rho_{1,P}P_xW_{\theta_1\theta_2}-\rho_{1,\theta_2}W_{x\theta_1}\\ &+\rho_{1,W}W_{\theta_2}W_{x\theta_1}+\rho_{1,P}P_{\theta_2}W_{x\theta_1}-\rho_{2,x\theta_2}W_{\theta_2}+\rho_{2,xP}P_{\theta_2}W_{\theta_2}-\rho_{2,\theta_2W}W_xW_{\theta_2}\\ &+\rho_{2,WP}P_{\theta_2}W_xW_{\theta_2}+\rho_{2,W}W_{x\theta_2}W_{\theta_2}-\rho_{2,\theta_2P}P_xW_{\theta_2}+\rho_{2,PP}P_xP_{\theta_2}W_{\theta_2}\\ &+\rho_{2,P}P_{x\theta_2}W_{\theta_2}-\rho_{2,\theta_2}W_{x\theta_2}+\rho_{2,W}W_{\theta_2}W_{x\theta_2}+\rho_{2,P}P_{\theta_2}W_{x\theta_2},
\end{split}
\label{bigcoeff3}
\end{equation}

\begin{equation}
\begin{split}
\Lambda^{\theta_1\theta_2}=&\Lambda_{\theta_1\theta_2}-\Lambda_{\theta_1W}W_{\theta_2}-\Lambda_{\theta_1P}P_{\theta_2}+\Lambda_{\theta_2W}W_{\theta_1}-\Lambda_{WW}W_{\theta_1}W_{\theta_2}-\Lambda_{WP}W_{\theta_1}P_{\theta_2}\\ &+\Lambda_{W}W_{\theta_1\theta_2}+\Lambda_{\theta_2P}P_{\theta_1}-\Lambda_{WP}P_{\theta_1}W_{\theta_2}-\Lambda_{PP}P_{\theta_1}P_{\theta_2}+\Lambda_{P}P_{\theta_1\theta_2}-\xi_{\theta_1\theta_2}W_x\\ &+\xi_{\theta_1W}W_xW_{\theta_2}+\xi_{\theta_1P}W_xP_{\theta_2}+\xi_{\theta_1}W_{x\theta_2}-\xi_{\theta_2W}W_xW_{\theta_1}+\xi_{WW}W_xW_{\theta_1}W_{\theta_2}\\ &+\xi_{WP}W_xW_{\theta_1}P_{\theta_2}+\xi_{W}W_{\theta_1}W_{x\theta_2}-\xi_{W}W_xW_{\theta_1\theta_2}-\xi_{\theta_2P}W_xP_{\theta_1}\\ &+\xi_{WP}P_{\theta_1}W_xW_{\theta_2}+\xi_{PP}W_xP_{\theta_1}P_{\theta_2}+\xi_{P}P_{\theta_1}W_{x\theta_2}-\xi_{P}W_xP_{\theta_1\theta_2}-\xi_{\theta_2}W_{x\theta_1}\\ &-\xi_{W}W_{\theta_2}W_{x\theta_1}-\xi_{P}P_{\theta_2}W_{x\theta_1}-\tau_{\theta_1\theta_2}W_t+\tau_{\theta_1W}W_tW_{\theta_2}+\tau_{\theta_1P}W_tP_{\theta_2}+\tau_{\theta_1}W_{t\theta_2}\\ &-\tau_{\theta_2W}W_tW_{\theta_1}+\tau_{WW}W_tW_{\theta_1}W_{\theta_2}+\tau_{WP}W_tW_{\theta_1}P_{\theta_2}+\tau_{W}W_{\theta_1}W_{t\theta_2}\\ &-\tau_{W}W_tW_{\theta_1\theta_2}-\tau_{\theta_2P}W_tP_{\theta_1}+\tau_{WP}P_{\theta_1}W_tW_{\theta_2}+\tau_{PP}W_tP_{\theta_1}P_{\theta_2}+\tau_{P}P_{\theta_1}W_{t\theta_2}\\ &-\tau_{P}W_tP_{\theta_1\theta_2}-\tau_{\theta_2}W_{t\theta_1}-\tau_{W}W_{\theta_2}W_{t\theta_1}-\tau_{P}P_{\theta_2}W_{t\theta_1}-\rho_{1,\theta_1\theta_2}W_{\theta_1}\\ &+\rho_{1,\theta_1W}W_{\theta_1}W_{\theta_2}+\rho_{1,\theta_1P}W_{\theta_1}P_{\theta_2}-\rho_{1,\theta_1}W_{\theta_1\theta_2}+\rho_{1,\theta_2P}P_{\theta_1}W_{\theta_1}\\ &-\rho_{1,WP}P_{\theta_1}W_{\theta_1}W_{\theta_2}+\rho_{1,PP}P_{\theta_1}P_{\theta_2}W_{\theta_1}-\rho_{1,P}W_{\theta_1}P_{\theta_1\theta_2}-\rho_{1,P}P_{\theta_1}W_{\theta_1\theta_2}\\ &-\rho_{2,\theta_1\theta_2}W_{\theta_2}-\rho_{2,\theta_1P}P_{\theta_2}W_{\theta_2}+\rho_{2,\theta_2W}W_{\theta_1}W_{\theta_2}-\rho_{2,WP}P_{\theta_2}W_{\theta_1}W_{\theta_2}\\ &+\rho_{2,\theta_2P}P_{\theta_1}W_{\theta_2}+\rho_{2,PP}P_{\theta_1}P_{\theta_2}W_{\theta_2}-\rho_{2,P}W_{\theta_2}P_{\theta_1\theta_2}-\rho_{2,\theta_2}W_{\theta_1\theta_2}+\rho_{2,P}P_{\theta_2}W_{\theta_1\theta_2},
\end{split}
\label{bigcoeff4}
\end{equation}

\begin{equation}
\begin{split}
\Omega^{xt}=&\Omega_{xt}+\Omega_{xP}P_t+\Omega_{xW}W_t+\Omega_{tP}P_x+\Omega_{PP}P_xP_t+\Omega_{PW}P_xW_t+\Omega_{P}P_{xt}\\ &+\Omega_{tW}W_x+\Omega_{PW}W_xP_t+\Omega_{WW}W_xW_t+\Omega_{W}W_{xt}-\xi_{xt}P_x-\xi_{xP}P_xP_t-\xi_{xW}P_xW_t\\ &-\xi_xP_{xt}-\xi_{tP}(P_x)^2-\xi_{PP}(P_x)^2P_t-\xi_{PW}(P_x)^2W_t-2\xi_{P}P_xP_{xt}\\ &-\xi_{tW}P_xW_x-\xi_{PW}P_xP_tW_x-\xi_{WW}P_xW_xW_t-\xi_{W}W_xP_{xt}-\xi_{W}P_xW_{xt}-\xi_tP_{xx}\\ &-\xi_{P}P_tP_{xx}-\xi_{W}W_tP_{xx}-\tau_{xt}P_t-\tau_{xP}(P_t)^2-\tau_{xW}P_tW_t-\tau_xP_{tt}-\tau_{tP}P_xP_t\\ &-\tau_{PP}(P_t)^2P_x-\tau_{PW}P_xP_tW_t-2\tau_{P}P_tP_{xt}-\tau_{P}P_xP_{tt}-\tau_{tW}P_tW_x\\ &-\tau_{PW}(P_t)^2W_x-\tau_{WW}P_tW_xW_t-\tau_{W}W_xP_{tt}-\tau_{W}P_tW_{xt}-\tau_tP_{xt}-\tau_{W}W_tP_{xt}\\ 
&-\rho_{1,xt}P_{\theta_1}-\rho_{1,xP}P_tP_{\theta_1}-\rho_{1,xW}W_tP_{\theta_1}-\rho_{1,tP}P_xP_{\theta_1}-\rho_{1,x}P_{t\theta_1}-\rho_{1,t}P_{x\theta_1}\\ &-\rho_{1,PP}P_xP_tP_{\theta_1}-\rho_{1,PW}P_xP_{\theta_1}W_t-\rho_{1,P}P_{xt}P_{\theta_1}-\rho_{1,P}P_{t\theta_1}P_x\\ &-\rho_{1,P}P_{x\theta_1}P_t-\rho_{1,tW}W_xP_{\theta_1}-\rho_{1,PW}W_xP_tP_{\theta_1}-\rho_{1,WW}W_xW_tP_{\theta_1}-\rho_{1,W}W_{xt}P_{\theta_1}\\ &-\rho_{1,W}W_xP_{t\theta_1}-\rho_{1,W}W_tP_{x\theta_1}-\rho_{2,xt}P_{\theta_2}-\rho_{2,xP}P_tP_{\theta_2}-\rho_{2,xW}W_tP_{\theta_2}\\ &-\rho_{2,tP}P_xP_{\theta_2}-\rho_{2,x}P_{t\theta_2}-\rho_{2,t}P_{x\theta_2}-\rho_{2,PP}P_xP_tP_{\theta_2}-\rho_{2,PW}W_tP_xP_{\theta_2}\\ &-\rho_{2,P}P_{xt}P_{\theta_2}-\rho_{2,P}P_{t\theta_2}P_x-\rho_{2,P}P_{x\theta_2}P_t-\rho_{2,tW}W_xP_{\theta_2}-\rho_{2,PW}W_xP_tP_{\theta_2}\\ &-\rho_{2,WW}W_xW_tP_{\theta_2}-\rho_{2,W}W_{xt}P_{\theta_2}-\rho_{2,W}P_{t\theta_2}W_x-\rho_{2,W}P_{x\theta_2}W_t,
\end{split}
\label{bigcoeff1b}
\end{equation}

\begin{equation}
\begin{split}
\Omega^{t\theta_1}=&\Omega_{t\theta_1}+\Omega_{tP}P_{\theta_1}+\Omega_{tW}W_{\theta_1}+\Omega_{\theta_1P}P_t+\Omega_{PP}P_tP_{\theta_1}+\Omega_{PW}P_tW_{\theta_1}+\Omega_{P}P_{t\theta_1}\\ &+\Omega_{\theta_1W}W_t+\Omega_{PW}W_tP_{\theta_1}+\Omega_{WW}W_tW_{\theta_1}+\Omega_{W}W_{t\theta_1}
-\xi_{t\theta_1}P_x-\xi_{tP}P_xP_{\theta_1}\\ &-\xi_{tW}P_xW_{\theta_1}-\xi_tP_{x\theta_1}-\xi_{\theta_1P}P_xP_t-\xi_{PP}P_xP_tP_{\theta_1}-\xi_{PW}P_xP_tW_{\theta_1}\\ &-\xi_{P}P_tP_{x\theta_1}-\xi_{P}P_xP_{t\theta_1}-\xi_{\theta_1W}P_xW_t-\xi_{PW}W_tP_xP_{\theta_1}-\xi_{WW}P_xW_tW_{\theta_1}\\ &-\xi_{W}W_tP_{x\theta_1}-\xi_{W}P_xW_{t\theta_1}-\xi_{\theta_1}P_{xt}-\xi_{P}P_{xt}P_{\theta_1}-\xi_{W}P_{xt}W_{\theta_1}-\tau_{t\theta_1}P_t\\ &-\tau_{tP}P_tP_{\theta_1}-\tau_{tW}P_tW_{\theta_1}-\tau_tP_{t\theta_1}-\tau_{\theta_1P}(P_t)^2-\tau_{PP}(P_t)^2P_{\theta_1}\\ &-\tau_{PW}(P_t)^2W_{\theta_1}-2\tau_{P}P_tP_{t\theta_1}-\tau_{\theta_1W}P_tW_t-\tau_{PW}W_tP_tP_{\theta_1}-\tau_{WW}P_tW_tW_{\theta_1}\\ &-\tau_{W}W_tP_{t\theta_1}-\tau_{W}P_tW_{t\theta_1}-\tau_{\theta_1}P_{tt}-\tau_{P}P_{tt}P_{\theta_1}-\tau_{W}P_{tt}W_{\theta_1}-\rho_{1,t\theta_1}P_{\theta_1}\\ &+\rho_{1,tW}W_{\theta_1}P_{\theta_1}-\rho_{1,\theta_1P}P_tP_{\theta_1}+\rho_{1,PW}W_{\theta_1}P_tP_{\theta_1}+\rho_{1,P}P_{t\theta_1}P_{\theta_1}\\ &-\rho_{1,\theta_1W}W_tP_{\theta_1}+\rho_{1,WW}W_tW_{\theta_1}P_{\theta_1}+\rho_{1,W}W_{t\theta_1}P_{\theta_1}-\rho_{1,\theta_1}P_{t\theta_1}+\rho_{1,P}P_{\theta_1}P_{t\theta_1}\\ &+\rho_{1,W}W_{\theta_1}P_{t\theta_1}-\rho_{2,t\theta_1}P_{\theta_2}+\rho_{2,tP}P_{\theta_1}P_{\theta_2}+\rho_{2,tW}W_{\theta_1}P_{\theta_2}-\rho_{2,t}P_{\theta_1\theta_2}\\ &-\rho_{2,\theta_1P}P_tP_{\theta_2}+\rho_{2,PP}P_tP_{\theta_1}P_{\theta_2}+\rho_{2,PW}W_{\theta_1}P_tP_{\theta_2}+\rho_{2,P}P_{t\theta_1}P_{\theta_2}\\ &-\rho_{2,P}P_tP_{\theta_1\theta_2}-\rho_{2,\theta_1W}W_tP_{\theta_2}+\rho_{2,PW}W_tP_{\theta_1}P_{\theta_2}+\rho_{2,WW}W_tW_{\theta_1}P_{\theta_2}\\ &+\rho_{2,W}W_{t\theta_1}P_{\theta_2}-\rho_{2,W}W_tP_{\theta_1\theta_2}-\rho_{2,\theta_1}P_{t\theta_2}+\rho_{2,P}P_{\theta_1}P_{t\theta_2}+\rho_{2,W}W_{\theta_1}P_{t\theta_2},
\end{split}
\label{bigcoeff2b}
\end{equation}

\begin{equation}
\begin{split}
\Omega^{x\theta_2}=&\Omega_{x\theta_2}+\Omega_{xP}P_{\theta_2}+\Omega_{xW}W_{\theta_2}+\Omega_{\theta_2P}P_x+\Omega_{PP}P_xP_{\theta_2}+\Omega_{PW}P_xW_{\theta_2}+\Omega_{P}P_{x\theta_2}\\ &+\Omega_{\theta_2W}W_x+\Omega_{PW}W_xP_{\theta_2}+\Omega_{WW}W_xW_{\theta_2}+\Omega_{W}W_{x\theta_2}-\xi_{x\theta_2}P_x-\xi_{xP}P_xP_{\theta_2}-\xi_{xW}P_xW_{\theta_2}\\ &-\xi_xP_{x\theta_2}-\xi_{\theta_2P}(P_x)^2-\xi_{PP}(P_x)^2P_{\theta_2}-\xi_{PW}(P_x)^2W_{\theta_2}-2\xi_{P}P_xP_{x\theta_2}\\ &-\xi_{\theta_2W}P_xW_x-\xi_{PW}W_xP_xP_{\theta_2}-\xi_{WW}P_xW_xW_{\theta_2}-\xi_{W}W_xP_{x\theta_2}-\xi_{W}P_xW_{x\theta_2}\\ &-\xi_{\theta_2}P_{xx}-\xi_{P}P_{xx}P_{\theta_2}-\xi_{W}P_{xx}W_{\theta_2}-\tau_{x\theta_2}P_t-\tau_{xP}P_tP_{\theta_2}-\tau_{xW}P_tW_{\theta_2}\\ &-\tau_xP_{t\theta_2}-\tau_{\theta_2P}P_xP_t-\tau_{PP}P_xP_tP_{\theta_2}-\tau_{PW}P_xP_tW_{\theta_2}-\tau_{P}P_xP_{t\theta_2}\\ &-\tau_{P}P_tP_{x\theta_2}-\tau_{\theta_2W}P_tW_x-\tau_{PW}W_xP_tP_{\theta_2}-\tau_{WW}P_tW_xW_{\theta_2}-\tau_{W}W_xP_{t\theta_2}\\ &-\tau_{W}P_tW_{x\theta_2}-\tau_{\theta_2}P_{xt}-\tau_{P}P_{xt}P_{\theta_2}-\tau_{W}P_{xt}W_{\theta_2}-\rho_{1,x\theta_2}P_{\theta_1}+\rho_{1,xP}P_{\theta_2}P_{\theta_1}\\ &+\rho_{1,xW}W_{\theta_2}P_{\theta_1}+\rho_{1,x}P_{\theta_1\theta_2}-\rho_{1,\theta_2P}P_xP_{\theta_1}+\rho_{1,PP}P_xP_{\theta_2}P_{\theta_1}\\ &+\rho_{1,PW}W_{\theta_2}P_xP_{\theta_1}+\rho_{1,P}P_{x\theta_2}P_{\theta_1}+\rho_{1,P}P_xP_{\theta_1\theta_2}-\rho_{1,\theta_2W}W_xP_{\theta_1}\\ &+\rho_{1,PW}W_xP_{\theta_2}P_{\theta_1}+\rho_{1,WW}W_xW_{\theta_2}P_{\theta_1}+\rho_{1,W}W_{x\theta_2}P_{\theta_1}+\rho_{1,W}W_xP_{\theta_1\theta_2}-\rho_{1,\theta_2}P_{x\theta_1}\\ &+\rho_{1,P}P_{\theta_2}P_{x\theta_1}+\rho_{1,W}W_{\theta_2}P_{x\theta_1}-\rho_{2,x\theta_2}P_{\theta_2}+\rho_{2,xW}W_{\theta_2}P_{\theta_2}-\rho_{2,\theta_2P}P_xP_{\theta_2}\\ &+\rho_{2,PW}W_{\theta_2}P_xP_{\theta_2}+\rho_{2,P}P_{x\theta_2}P_{\theta_2}-\rho_{2,\theta_2W}W_xP_{\theta_2}+\rho_{2,WW}W_xW_{\theta_2}P_{\theta_2}\\ &+\rho_{2,W}W_{x\theta_2}P_{\theta_2}-\rho_{2,\theta_2}P_{x\theta_2}+\rho_{2,P}P_{\theta_2}P_{x\theta_2}+\rho_{2,W}W_{\theta_2}P_{x\theta_2},
\end{split}
\label{bigcoeff3b}
\end{equation}

and

\begin{equation}
\begin{split}
\Omega^{\theta_1\theta_2}=&\Omega_{\theta_1\theta_2}-\Omega_{\theta_1P}P_{\theta_2}-\Omega_{\theta_1W}W_{\theta_2}+\Omega_{\theta_2P}P_{\theta_1}-\Omega_{PP}P_{\theta_1}P_{\theta_2}-\Omega_{PW}P_{\theta_1}W_{\theta_2}\\ &+\Omega_{P}P_{\theta_1\theta_2}+\Omega_{\theta_2W}W_{\theta_1}-\Omega_{PW}W_{\theta_1}P_{\theta_2}-\Omega_{WW}W_{\theta_1}W_{\theta_2}+\Omega_{W}W_{\theta_1\theta_2}-\xi_{\theta_1\theta_2}P_x\\ &+\xi_{\theta_1P}P_xP_{\theta_2}+\xi_{\theta_1W}P_xW_{\theta_2}+\xi_{\theta_1}P_{x\theta_2}-\xi_{\theta_2P}P_xP_{\theta_1}+\xi_{PP}P_xP_{\theta_1}P_{\theta_2}\\ &+\xi_{PW}P_xP_{\theta_1}W_{\theta_2}+\xi_{P}P_{\theta_1}P_{x\theta_2}-\xi_{P}P_xP_{\theta_1\theta_2}-\xi_{\theta_2W}P_xW_{\theta_1}\\ &+\xi_{PW}W_{\theta_1}P_xP_{\theta_2}+\xi_{WW}P_xW_{\theta_1}W_{\theta_2}+\xi_{W}W_{\theta_1}P_{x\theta_2}-\xi_{W}P_xW_{\theta_1\theta_2}-\xi_{\theta_2}P_{x\theta_1}\\ &-\xi_{P}P_{\theta_2}P_{x\theta_1}-\xi_{W}W_{\theta_2}P_{x\theta_1}-\tau_{\theta_1\theta_2}P_t+\tau_{\theta_1P}P_tP_{\theta_2}+\tau_{\theta_1W}P_tW_{\theta_2}+\tau_{\theta_1}P_{t\theta_2}\\ &-\tau_{\theta_2P}P_tP_{\theta_1}+\tau_{PP}P_tP_{\theta_1}P_{\theta_2}+\tau_{PW}P_tP_{\theta_1}W_{\theta_2}+\tau_{P}P_{\theta_1}P_{t\theta_2}\\ &-\tau_{P}P_tP_{\theta_1\theta_2}-\tau_{\theta_2W}P_tW_{\theta_1}+\tau_{PW}W_{\theta_1}P_tP_{\theta_2}+\tau_{WW}P_tW_{\theta_1}W_{\theta_2}+\tau_{W}W_{\theta_1}P_{t\theta_2}\\ &-\tau_{W}P_tW_{\theta_1\theta_2}-\tau_{\theta_2}P_{t\theta_1}-\tau_{P}P_{\theta_2}P_{t\theta_1}-\tau_{W}W_{\theta_2}P_{t\theta_1}-\rho_{1,\theta_1\theta_2}P_{\theta_1}\\ &+\rho_{1,\theta_1P}P_{\theta_1}P_{\theta_2}+\rho_{1,\theta_1W}P_{\theta_1}W_{\theta_2}-\rho_{1,\theta_1}P_{\theta_1\theta_2}+\rho_{1,\theta_2W}W_{\theta_1}P_{\theta_1}\\ &-\rho_{1,PW}W_{\theta_1}P_{\theta_1}P_{\theta_2}+\rho_{1,WW}W_{\theta_1}W_{\theta_2}P_{\theta_1}-\rho_{1,W}P_{\theta_1}W_{\theta_1\theta_2}-\rho_{1,W}W_{\theta_1}P_{\theta_1\theta_2}\\ &-\rho_{2,\theta_1\theta_2}P_{\theta_2}-\rho_{2,\theta_1W}W_{\theta_2}P_{\theta_2}+\rho_{2,\theta_2P}P_{\theta_1}P_{\theta_2}-\rho_{2,PW}W_{\theta_2}P_{\theta_1}P_{\theta_2}\\ &+\rho_{2,\theta_2W}W_{\theta_1}P_{\theta_2}+\rho_{2,WW}W_{\theta_1}W_{\theta_2}P_{\theta_2}-\rho_{2,W}P_{\theta_2}W_{\theta_1\theta_2}-\rho_{2,\theta_2}P_{\theta_1\theta_2}+\rho_{2,W}W_{\theta_2}P_{\theta_1\theta_2},
\end{split}
\label{bigcoeff4b}
\end{equation}

Substituting the above formulae into the equations (\ref{cccccc1}) and (\ref{cccccc2}) and making the substitutions from the original supersymmetric equations (\ref{c1})
\begin{equation}
\begin{split}
&P_t = P_x(D_1D_2W)+(D_1D_2P)W_x,\\
& \\
&W_t = (D_1D_2W)W_x+(-1)^{\gamma}A(D_1D_2P)^{\gamma-2}(D_1P)(D_2P)(D_1D_2P)_x,
\end{split}
\label{substthishere1}
\end{equation}
we obtain a series of determining equations for the functions $\xi$, $\tau$, $\rho_1$, $\rho_2$, $\Lambda$ and $\Omega$ in (\ref{symmie3}). We provide a specific solution of these determining equations, with the form:
\begin{equation}
\begin{split}
\xi(x,\theta_1)&=2C_1x+C_2-\underline{K_1}\theta_1,\qquad
\tau(t,\theta_2)=2C_3t+C_4-\underline{K_2}\theta_2,\\
\rho_1(\theta_1)&=C_1\theta_1+\underline{K_1},\qquad
\rho_2(\theta_2)=C_3\theta_2+\underline{K_2},\qquad
\Lambda(W)=(3C_1-C_3)W,\\
\Omega(P,\gamma)&=\left({(\gamma+5)C_1+(\gamma-3)C_3\over (\gamma+1)}\right)P
\end{split}
\label{symmie8}
\end{equation}
where $C_1$, $C_2$, $C_3$ are bosonic constants, while $\underline{K_1}$ and $\underline{K_2}$ are fermionic constants. Here, the function $\Omega(P,\gamma)$ is parametrized by the polytropic exponent $\gamma$. Thus, the superalgebra $\mathfrak{L}$ is spanned by the following vector fields:

\begin{equation}
\begin{split}
& P_1=\partial_x,\qquad P_2=\partial_t,\qquad Q_1=-\theta_1\partial_x+\partial_{\theta_1},\qquad Q_2=-\theta_2\partial_t+\partial_{\theta_2},\\
& L_1(\gamma)=2x\partial_x+\theta_1\partial_{\theta_1}+3W\partial_W+\left(\frac{\gamma+5}{\gamma+1}\right)P\partial_P,\\ & L_2(\gamma)=2t\partial_t+\theta_2\partial_{\theta_2}-W\partial_W+\left(\frac{\gamma-3}{\gamma+1}\right)P\partial_P.
\end{split}
\label{h1}
\end{equation}
Here, the generators $P_1$ and $P_2$ represent translations in the $x$ and $t$ directions respectively, $L_1(\gamma)$ and $L_2(\gamma)$ are two families of dilations in the independent and dependent variables, and $Q_1$ and $Q_2$ are the supersymmetric transformations already determined in equation (\ref{b3}).
The dependence on $\gamma$ is reflected in the form of the two dilations, $L_1(\gamma)$ and $L_2(\gamma)$, which we abbreviate for convenience as $L_1$ and $L_2$ respectively. The supercommutation relations of the superalgebra generators (\ref{h1}) are given in Table I.

\begin{table}[htbp]
  \begin{center}
\caption{Supercommutation table for the Lie superalgebra $\mathfrak{L}$ spanned by the
  vector fields (\ref{h1})}
\vspace{5mm}
\setlength{\extrarowheight}{4pt}
\begin{tabular}{|c||c|c|c|c|c|c|}\hline
& $\mathbf{L_1}$ & $\mathbf{P_1}$ & $\mathbf{Q_1}$ & $\mathbf{L_2}$ & $\mathbf{P_2}$ & $\mathbf{Q_2}$\\[0.5ex]\hline\hline
$\mathbf{L_1}$ & $0$ & $-2P_1$ & $-Q_1$ & $0$ & $0$ & $0$ \\\hline
$\mathbf{P_1}$ & $2P_1$ & $0$ & $0$ & $0$ & $0$ & $0$ \\\hline
$\mathbf{Q_1}$ & $Q_1$ & $0$ & $-2P_1$ & $0$ & $0$ & $0$ \\\hline
$\mathbf{L_2}$ & $0$ & $0$ & $0$ & $0$ & $-2P_2$ & $-Q_2$ \\\hline
$\mathbf{P_2}$ & $0$ & $0$ & $0$ & $2P_2$ & $0$ & $0$ \\\hline
$\mathbf{Q_2}$ & $0$ & $0$ & $0$ & $Q_2$ & $0$ & $-2P_2$ \\\hline
\end{tabular}
  \end{center}
\end{table}

This superalgebra differs in structure from a classical Lie algebra in the sense that the diagonal contains non-zero elements, i.e. $\{Q_1,Q_1\}=-2P_1$ and $\{Q_2,Q_2\}=-2P_2$.

We now proceed to classify the one-dimensional subalgebras of the Lie superalgebra into conjugacy classes according to the action by the Lie supergroup generated by the Lie superalgebra $\mathfrak{L}$. We construct a list of representative subalgebras such that each subalgebra of $\mathfrak{L}$ is conjugate to one and only one element of the list. We perform this analysis using the methods described in (for example) \cite{Patera,Winternitz1,Winternitz2,Kac}.
We begin our classification by writing the superalgebra $\mathfrak{L}$ as a direct sum of two semi-direct sums of smaller algebras in the following way
\begin{equation}
\mathfrak{L}=\Big{\{}\{L_1\}\sdir\{P_1,Q_1\}\Big{\}}\oplus\Big{\{}\{L_2\}\sdir\{P_2,Q_2\}\Big{\}},
\label{j1}
\end{equation}

Consider first the semi-direct sum
\begin{equation}
\mathfrak{a}=\{L_1\}\sdir\{P_1,Q_1\}.
\end{equation}
This algebra $\mathfrak{a}$ can be classified by using the method for semi-direct sums of algebras \cite{Winternitz1}. This method leads to splitting subalgebras (of the form $\mathfrak{m}\sdir\mathfrak{n}$, where $\mathfrak{m}\subset\{L_1\}$ and $\mathfrak{n}\subset\{P_1,Q_1\}$) and non-splitting subalgebras which cannot be written in this form.

Let us first discuss splitting subalgebras. The subalgebras of $\{L_1\}$ are $\{0\}$ and $\{L_1\}$. For each of these two subalgebras, we must consider all subalgebras of $\{P_1,Q_1\}$ which are invariant under (super)commutation under the given subalgebra of $\{L_1\}$.

For subalgebra $\{0\}$, all subspaces of $\{P_1,Q_1\}$ are invariant subalgebras. That is,
\begin{equation}
[\alpha P_1+\underline{\mu}Q_1,0]=0\subset\{0\}
\end{equation}
So the possible representative subalgebras are $\{0\}$, $\{P_1\}$, $\{\underline{\mu}Q_1\}$, $\{P_1+\underline{\mu}Q_1\}$ and $\{P_1,Q_1\}$

We classify such subalgebras in the following way. We begin with the trivial subalgebra $\mathfrak{a}_0=\{0\}$. Let us now consider the one-dimensional subalgebra $\{P_1\}$
and apply the action of the subgroup $e^{\{L_1,P_1,Q_1\}}$ generated by $\mathfrak{a}$ to it. If we take any arbitrary element of $\mathfrak{a}$:
\begin{equation}
Y=\alpha L_1+\beta P_1+\underline{\eta} Q_1,
\label{neweq2}
\end{equation}
then
\begin{equation}
[Y,P_1]=-2\alpha P_1.
\label{neweq3}
\end{equation}
Therefore, under the action of the subgroup $e^{\{L_1,P_1,Q_1\}}$, we obtain $P_1\rightarrow e^{-2\alpha}P_1$. So, $\mathfrak{a}_1=\{P_1\}$ is the only subalgebra in its conjugation class.

Consider now the subalgebra $\{\underline{\mu}Q_1\}$. If $Y=\alpha L_1+\beta P_1+\underline{\eta} Q_1$, then we obtain
\begin{equation}
[Y,\underline{\mu}Q_1]=-\alpha\underline{\mu}Q_1+\underline{\eta}\underline{\mu}P_1.
\label{neweq5}
\end{equation}
Consequently, under the group action of $e^{\{L_1,P_1,Q_1\}}$, we get
\begin{equation}
\{\underline{\mu}Q_1\}\rightarrow\begin{cases}\underline{\mu}Q_1+{1\over \alpha}\left(1-e^{-\alpha}\right)\underline{\eta}\underline{\mu}P_1,\qquad\mbox{if }\alpha\neq 0\\ \underline{\mu}Q_1+\underline{\eta}\underline{\mu}P_1,\qquad\mbox{if }\alpha=0\end{cases}
\label{neweq6}
\end{equation}
So, by re-scaling the $\underline{\eta}$, we can obtain other multiples of the generator $\underline{\mu}Q_1+\underline{\eta}\underline{\mu}P_1$.
We represent all these possibilities by the subalgebra $\mathfrak{a}_2=\{\underline{\mu}Q_1\}$.

The subalgebra $\{P_1+\underline{\mu}Q_1\}$, under the action of the group, leads to
\begin{equation}
\{P_1+\underline{\mu}Q_1\}\rightarrow \{\underline{\mu}Q_1+(1+\underline{\eta}\underline{\mu})P_1\}.
\end{equation}
All of these are covered under the subalgebra $\mathfrak{a}_3=\{P_1+\underline{\mu}Q_1\}$.

The remaining subalgebra, $\{P_1,Q_1\}$ is already two-dimensional and is therefore not included in our classification of one-dimensional subalgebras of $\mathfrak{a}$.

The subalgebra $\{L_1\}$ is already one-dimensional, so the only splitting one-dimensional subalgebra of $\mathfrak{a}$ obtained from consideration of the subalgebras of $\{P_1,Q_1\}$ invariant under $\{L_1\}$ is $\mathfrak{a}_4=\{L_1\}$ itself.

Let us now discuss non-splitting subalgebras of $\mathfrak{a}$. We consider vector spaces of the form
\begin{equation}
V=\{L_1+a_1P_1+\underline{a_2}Q_1\}.
\label{theeq57}
\end{equation}
Applying a cocycle to $\{L_1\}$, we obtain
\begin{equation}
L_1+\lambda_1[P_1,L_1]+\underline{\lambda_2}[Q_1,L_1] = L_1+2\lambda_1P_1+\underline{\lambda_2}Q_1
\end{equation}
so, if we let $\lambda_1=-\frac{1}{2}a_1$ and $\underline{\lambda_2}=-\underline{a_2}$, then for (\ref{theeq57}), we get
\begin{equation}
V\rightarrow\{L_1+2\lambda_1P_1+\underline{\lambda_2}Q_1+c_1P_1+\underline{D_1}Q_1\}=\{L_1\}
\end{equation}
Thus, there are no non-splitting subalgebras of $\mathfrak{a}$.
So, the algebra $\mathfrak{a}=\{L_1,P_1,Q_1\}$ has the one-dimensional subalgebra classification
\begin{equation}
\mathfrak{a}_0=\{0\},\qquad \mathfrak{a}_1=\{P_1\},\qquad \mathfrak{a}_2=\{\underline{\mu}Q_1\},\qquad \mathfrak{a}_3=\{P_1+\underline{\mu}Q_1\},\qquad \mathfrak{a}_4=\{L_1\}
\end{equation}
Similarily, the algebra $\mathfrak{b}=\{L_2,P_2,Q_2\}$ has the one-dimensional subalgebra classification
\begin{equation}
\mathfrak{b}_0=\{0\},\qquad \mathfrak{b}_1=\{P_2\},\qquad \mathfrak{b}_2=\{\underline{\mu}Q_2\},\qquad \mathfrak{b}_3=\{P_2+\underline{\mu}Q_2\},\qquad \mathfrak{b}_4=\{L_2\}.
\end{equation}

Considering now the direct sum
\begin{equation}
\mathfrak{L}=\mathfrak{a}\oplus\mathfrak{b}=\{L_1,P_1,Q_1\}\oplus\{L_2,P_2,Q_2\}
\label{newquestion1}
\end{equation}
we use the Goursat method for the classification of direct sums \cite{Goursat,DuVal}.

The one-dimensional non-twisted subalgebras are formulated by successively taking the direct sums of each subalgebra of $\mathfrak{a}$ with each subalgebra of $\mathfrak{b}$. This results in the following list:
\begin{equation}
\begin{split}
&\mathfrak{L}_1=\{P_1\},\qquad \mathfrak{L}_2=\{\underline{\mu}Q_1\},\qquad \mathfrak{L}_3=\{P_1+\underline{\mu}Q_1\},\qquad \mathfrak{L}_4=\{L_1\},\\ & \mathfrak{L}_5=\{P_2\},\qquad \mathfrak{L}_6=\{\underline{\nu}Q_2\},\qquad \mathfrak{L}_7=\{P_2+\underline{\nu}Q_2\},\qquad \mathfrak{L}_8=\{L_2\},\qquad 
\end{split}
\label{newquestion2}
\end{equation}

For twisted subalgebras, there are 16 possibilities of combining two subalgebras together. Two subalgebras, $\mathfrak{a}_i$ and $\mathfrak{b}_j$ can be twisted together if a homomorphism $\tau$ exists from $\mathfrak{a}_i$ to $\mathfrak{b}_j$:
\begin{equation}
\tau(\mathfrak{a}_i)=\mathfrak{b}_j.
\label{werewolf1}
\end{equation}
For example, if we consider $\mathfrak{a}_1=\{P_1\}$ and $\mathfrak{b}_1=\{P_2\}$, then the most general homomorphism from $\mathfrak{a}_1$ to $\mathfrak{b}_1$ is:
\begin{equation}
\tau:P_1\rightarrow kP_2,
\label{homomorphismindeed}
\end{equation}
where $k$ is any scalar. This gives us the family of twisted subalgebras $V=\{P_1+kP_2\}$. By considering the action of the full supergroup $e^{\{L_1,L_2,P_1,P_2,Q_1,Q_2\}}$ on $V$, we see that $k$ can be rescaled by any positive real number. So, the representative subalgebra is:
\begin{equation}
\mathfrak{L}_9=\{P_1+\varepsilon P_2\},\qquad \varepsilon=\pm 1.
\end{equation}
Considering all possibilities, we obtain the list of all representative one-dimensional subalgebras of the superalgebra $\mathfrak{L}$ generated by the vector fields (\ref{h1}):
\begin{eqnarray}
\mathfrak{L}_1=\{P_1\},&\nonumber
 &\mathfrak{L}_2=\{\underline{\mu}Q_1\},\\ \nonumber
 \mathfrak{L}_3=\{P_1+\underline{\mu}Q_1\},&\nonumber
&\mathfrak{L}_4=\{L_1\},\\ \nonumber
\mathfrak{L}_5=\{P_2\},&\nonumber
&\mathfrak{L}_6=\{\underline{\nu}Q_2\},\\ \nonumber
\mathfrak{L}_7=\{P_2+\underline{\nu}Q_2\},& \nonumber
&\mathfrak{L}_8=\{L_2\},\\ \nonumber
\mathfrak{L}_9=\{P_1+\varepsilon P_2,\varepsilon=\pm 1\},&\nonumber
&\mathfrak{L}_{10}=\{P_1+\underline{\nu}Q_2\},\\ \nonumber
\mathfrak{L}_{11}=\{P_1+\varepsilon P_2+\underline{\nu}Q_2,\varepsilon=\pm 1\},&\nonumber
&\mathfrak{L}_{12}=\{L_2+\varepsilon P_1,\varepsilon=\pm 1\},\\ \nonumber
\mathfrak{L}_{13}=\{P_2+\underline{\mu}Q_1\},&\nonumber
&\mathfrak{L}_{14}=\{\underline{\mu}Q_1+\underline{\nu}Q_2\},\\ \nonumber
\mathfrak{L}_{15}=\{P_2+\underline{\mu}Q_1+\underline{\nu}Q_2\},& \nonumber
&\mathfrak{L}_{16}=\{L_2+\underline{\mu}Q_1\},\\ \nonumber
\mathfrak{L}_{17}=\{P_1+\varepsilon P_2+\underline{\mu}Q_1,\varepsilon=\pm 1\}, &\nonumber
&\mathfrak{L}_{18}=\{P_1+\underline{\mu}Q_1+\underline{\nu}Q_2\},\\ \nonumber
\mathfrak{L}_{19}=\{P_1+\varepsilon P_2+\underline{\mu}Q_1+\underline{\nu}Q_2,\varepsilon=\pm 1\},&\nonumber
&\mathfrak{L}_{20}=\{L_2+\varepsilon P_1+\underline{\mu}Q_1,\varepsilon=\pm 1\},\\ \nonumber
\mathfrak{L}_{21}=\{L_1+\varepsilon P_2,\varepsilon=\pm 1\},&\nonumber
&\mathfrak{L}_{22}=\{L_1+\underline{\nu}Q_2\},\\ \nonumber
\mathfrak{L}_{23}=\{L_1+\varepsilon P_2+\underline{\nu}Q_2,\varepsilon=\pm 1\},&\nonumber
&\mathfrak{L}_{24}=\{L_1+kL_2,k\neq 0\}. \nonumber
\end{eqnarray}
This list of subalgebras will allow us to systematically use the symmetry reduction method \cite{Clarkson}.

\section{Symmetry reductions and invariant solutions}

We now make use of the symmetry reduction method in order to find invariant solutions of the supersymmetric polytropic gas dynamics equations (\ref{c1}). Replacing the specific forms of the superfields $W$ and $P$ as listed in Table II into the supersymmetric equation (\ref{c1}) we obtain symmetry reductions.

\begin{table}[htbp]
  \begin{center}
\caption{Invariants and change of variable for subalgebras of the Lie superalgebra $\mathfrak{L}$ spanned by the
  vector fields (\ref{h1})}
\vspace{3mm}
\setlength{\extrarowheight}{4pt}
\begin{tabular}{|c|c|c|}\hline
Subalgebra & Invariants & Superfields\\[0.5ex]\hline\hline

$\mathfrak{L}_1=\{P_1\}$ & $t$, $\theta_1$, $\theta_2$, $W$, $P$  & $W=W(t,\theta_1,\theta_2)$, $P=P(t,\theta_1,\theta_2)$ \\\hline

$\mathfrak{L}_3=\{P_1+\underline{\mu}Q_1\}$ & $t$, $\tau_1=\theta_1-\underline{\mu}x$, $\theta_2$, $W$, $P$ & $W=W(t,\tau_1,\theta_2)$, $P=P(t,\tau_1,\theta_2)$ \\\hline

$\mathfrak{L}_4=\{L_1\}$ & $t$, $\tau_1=x^{-1/2}\theta_1$, $\theta_2$, & $W=x^{3/2}{\mathcal A}\left(t,\tau_1,\theta_2\right)$, \\
& ${\mathcal A}=x^{-3/2}W$, ${\mathcal B}=x^{-\left({\gamma+5\over 2\gamma+2}\right)}P$ & $P=x^{\left({\gamma+5\over 2\gamma+2}\right)}{\mathcal B}\left(t,\tau_1,\theta_2\right)$\\\hline

$\mathfrak{L}_5=\{P_2\}$ & $x$, $\theta_1$, $\theta_2$, $W$, $P$  &  $W=W(x,\theta_1,\theta_2)$, $P=P(x,\theta_1,\theta_2)$\\\hline

$\mathfrak{L}_7=\{P_2+\underline{\nu}Q_2\}$ & $x$, $\theta_1$, $\tau_2=\theta_2-\underline{\nu}t$, $W$, $P$  & $W=W(x,\theta_1,\tau_2)$, $P=P(x,\theta_1,\tau_2)$ \\\hline

$\mathfrak{L}_8=\{L_2\}$ & $x$, $\theta_1$, $\tau_2=t^{-1/2}\theta_2$, & $W=t^{-1/2}{\mathcal A}\left(x,\theta_1,\tau_2\right)$, \\
& ${\mathcal A}=t^{1/2}W$, ${\mathcal B}=x^{-\left({\gamma-3\over 2\gamma+2}\right)}P$ & $P=x^{\left({\gamma-3\over 2\gamma+2}\right)}{\mathcal B}\left(t,\theta_1,\tau_2\right)$\\\hline

$\mathfrak{L}_9=\{P_1+\varepsilon P_2\}$ & $\sigma=x-\varepsilon t$, $\theta_1$, $\theta_2$, $W$, $P$  & $W=W(\sigma,\theta_1,\theta_2)$, $P=P(\sigma,\theta_1,\theta_2)$ \\\hline

$\mathfrak{L}_{10}=\{P_1+\underline{\nu}Q_2\}$ & $\sigma=t+\underline{\nu}\theta_2x$, $\theta_1$, & $W=W(\sigma,\theta_1,\tau_2)$, $P=P(\sigma,\theta_1,\tau_2)$ \\
 & $\tau_2=\theta_2-\underline{\nu}x$, $W$, $P$ & \\\hline

$\mathfrak{L}_{11}=\{P_1+\varepsilon P_2+\underline{\nu}Q_2\}$ & $\sigma=t-\varepsilon x+\underline{\nu}x\theta_2$, $\theta_1$, & $W=W(\sigma,\theta_1,\tau_2)$, $P=P(\sigma,\theta_1,\tau_2)$ \\
 &  $\tau_2=\theta_2-\underline{\nu}x$, $W$, $P$ &\\\hline

$\mathfrak{L}_{12}=\{L_2+\varepsilon P_1\}$ & $\sigma=x-\frac{1}{2}\varepsilon\ln{t}$, $\theta_1$, $\tau_2=t^{-1/2}\theta_2$, & $W=t^{-1/2}{\mathcal A}\left(\sigma,\theta_1,\tau_2\right)$ \\
& ${\mathcal A}=t^{1/2}W$, ${\mathcal B}=t^{-\left({\gamma-3\over 2\gamma+2}\right)}P$ & $P=t^{\left({\gamma-3\over 2\gamma+2}\right)}{\mathcal B}\left(\sigma,\theta_1,\tau_2\right)$\\\hline

$\mathfrak{L}_{13}=\{P_2+\underline{\mu}Q_1\}$ & $\sigma=x+\underline{\mu}\theta_1t$, $\tau_1=\theta_1-\underline{\mu}t$, & $W=W(\sigma,\tau_1,\theta_2)$ \\
 & $\theta_2$, $W$, $P$ & $P=P(\sigma,\tau_1,\theta_2)$ \\\hline

$\mathfrak{L}_{16}=\{L_2+\underline{\mu}Q_1\}$ & $\sigma=x+\frac{1}{2}\underline{\mu}\theta_1\ln{t}$, $\tau_1=\theta_1-\frac{1}{2}\underline{\mu}\ln{t}$, & $W=t^{-1/2}{\mathcal A}\left(\sigma,\tau_1,\tau_2\right)$ \\
& $\tau_2=t^{-1/2}\theta_2$, ${\mathcal A}=t^{1/2}W$, & $P=t^{\left({\gamma-3\over 2\gamma+2}\right)}{\mathcal B}\left(\sigma,\tau_1,\tau_2\right)$\\
& ${\mathcal B}=t^{-\left({\gamma-3\over 2\gamma+2}\right)}P$ & \\\hline

$\mathfrak{L}_{17}=\{P_1+\varepsilon P_2+\underline{\mu}Q_1\}$ & $\sigma=\varepsilon x-t+\underline{\mu}t\theta_1$, & $W=W(\sigma,\tau_1,\theta_2)$ \\
& $\tau_1=\theta_1-\varepsilon\underline{\mu}t$, $\theta_2$, $W$, $P$ & $P=P(\sigma,\tau_1,\theta_2)$ \\\hline

$\mathfrak{L}_{20}=\{L_2+\varepsilon P_1+\underline{\mu}Q_1\}$ & $\sigma=x+\frac{1}{2}\underline{\mu}\theta_1\ln{t}-\frac{1}{2}\varepsilon\ln{t}$,  & $W=t^{-1/2}{\mathcal A}(\sigma,\tau_1,\tau_2)$ \\
& $\tau_1=\theta_1-\frac{1}{2}\underline{\mu}\ln{t}$, & $P=t^{\left({\gamma-3\over 2\gamma+2}\right)}{\mathcal B}(\sigma,\tau_1,\tau_2)$\\
& $\tau_2=t^{-1/2}\theta_2$, ${\mathcal A}=t^{1/2}W$, & \\
& ${\mathcal B}=t^{-\left({\gamma-3\over 2\gamma+2}\right)}P$ & \\\hline

$\mathfrak{L}_{21}=\{L_1+\varepsilon P_2\}$ & $\sigma=t-\frac{1}{2}\varepsilon\ln{x}$, $\tau_1=x^{-1/2}\theta_1$, & $W=x^{3/2}{\mathcal A}\left(\sigma,\tau_1,\theta_2\right)$, \\
& $\theta_2$, ${\mathcal A}=x^{-3/2}W$, & $P=x^{\left({\gamma+5\over 2\gamma+2}\right)}{\mathcal B}\left(\sigma,\tau_1,\theta_2\right)$\\
& ${\mathcal B}=x^{-\left({\gamma+5\over 2\gamma+2}\right)}P$ &\\\hline

$\mathfrak{L}_{22}=\{L_1+\underline{\nu}Q_2\}$ & $\sigma=t+\frac{1}{2}\underline{\nu}\theta_2\ln{x}$, $\tau_1=x^{-1/2}\theta_1$, & $W=x^{3/2}{\mathcal A}\left(\sigma,\tau_1,\tau_2\right)$ \\
& $\tau_2=\theta_2-\frac{1}{2}\underline{\nu}\ln{x}$, ${\mathcal A}=x^{-3/2}W$, & $P=x^{\left({\gamma+5\over 2\gamma+2}\right)}{\mathcal B}\left(\sigma,\tau_1,\tau_2\right)$\\
& ${\mathcal B}=x^{-\left({\gamma+5\over 2\gamma+2}\right)}P$ &\\\hline

$\mathfrak{L}_{23}=\{L_1+\varepsilon P_2+\underline{\nu}Q_2\}$ & $\sigma=t+\frac{1}{2}\underline{\nu}\theta_2\ln{x}-\frac{1}{2}\varepsilon\ln{x}$,  & $W=x^{3/2}{\mathcal A}\left(\sigma,\tau_1,\tau_2\right)$ \\
& $\tau_1=x^{-1/2}\theta_1$, & $P=x^{\left({\gamma+5\over 2\gamma+2}\right)}{\mathcal B}\left(\sigma,\tau_1,\tau_2\right)$ \\
& $\tau_2=\theta_2-\frac{1}{2}\underline{\nu}\ln{x}$, ${\mathcal A}=x^{-3/2}W$, & \\
& ${\mathcal B}=x^{-\left({\gamma+5\over 2\gamma+2}\right)}P$ &\\\hline

$\mathfrak{L}_{24}=\{L_1+kL_2\}$ & $\sigma=t^{-\frac{1}{k}}x$, $\tau_1=t^{-\frac{1}{2k}}\theta_1$, & $W=t^{\frac{3-k}{2k}}{\mathcal A}\left(\sigma,\tau_1,\tau_2\right)$ \\
& $\tau_2=t^{-\frac{1}{2}}\theta_2$, ${\mathcal A}=t^{\frac{k-3}{2k}}W$, & $P=t^{\left(\frac{(\gamma+5)+k(\gamma-3)}{2k(\gamma+1)}\right)}{\mathcal B}\left(\sigma,\tau_1,\tau_2\right)$\\
& ${\mathcal B}=t^{-\left(\frac{(\gamma+5)+k(\gamma-3)}{2k(\gamma+1)}\right)}P$ &\\\hline

\end{tabular}
  \end{center}
\end{table}

In most cases these reductions involve heavy computations and we omit them for this reason. So, we present in this paper only certain classes of the invariant solutions involving (in most cases) freedom of arbitrary functions of one or two variables whose arguments can be either bosonic or fermionic. We propose two types of solutions. The first type has the form
\begin{equation}
W=S\left(\xi,F(\theta_1,\theta_2)\right),\qquad P=R\left(\xi,G(\theta_1,\theta_2)\right),
\end{equation}
where $F$ and $G$ are arbitrary functions of the fermionic arguments $\theta_1$ and $\theta_2$ and $\xi$ is the symmetry variable. In some cases, we specify them to be in the form
\begin{equation}
F(\theta_1,\theta_2)=\underline{\alpha}F_1(\theta_1)+\underline{\beta}F_2(\theta_2)\mbox{ and }G(\theta_1,\theta_2)=\underline{\kappa}G_1(\theta_1)+\underline{\lambda}G_2(\theta_2),
\end{equation}
where $\underline{\alpha}$, $\underline{\beta}$, $\underline{\kappa}$ and $\underline{\lambda}$ are fermionic constants.
The first type of solution is of the form
\begin{equation}
W=S\left(\xi,F(\theta_1,\theta_2)\right),\qquad P=R\left(\xi,G(\theta_1,\theta_2)\right),
\end{equation}
and corresponds to the results which we have obtained for subalgebras $\mathfrak{L}_1$, $\mathfrak{L}_3$, $\mathfrak{L}_7$, $\mathfrak{L}_8$, $\mathfrak{L}_{10}$, $\mathfrak{L}_{11}$, $\mathfrak{L}_{12}$, $\mathfrak{L}_{13}$, $\mathfrak{L}_{16}$, $\mathfrak{L}_{17}$, $\mathfrak{L}_{20}$, $\mathfrak{L}_{21}$, $\mathfrak{L}_{23}$ and $\mathfrak{L}_{24}$.

The second class of solution is of the type
\begin{equation}
W=S\left(\xi,\theta_1,\theta_2\right),\qquad P=R\left(\xi,\theta_1,\theta_2\right),
\label{thisonenow1}
\end{equation}
which we develop as a Taylor expansion with respect to $\theta_1$ and $\theta_2$ with non-constant coefficients which depend on the symmetry variable $\xi$. Since all terms that include $(\theta_1)^2$ or $(\theta_2)^2$ vanish, the solutions (\ref{thisonenow1}) are of the form
\begin{equation}
\begin{split}
&W=M_1(\xi)+\theta_1M_2(\xi)+\theta_2M_3(\xi)+\theta_1\theta_2M_4(\xi),\\ &P=N_1(\xi)+\theta_1N_2(\xi)+\theta_2N_3(\xi)+\theta_1\theta_2N_4(\xi)
\end{split}
\end{equation}
Replacing into the reduced equations, we obtain the solutions corresponding to subalgebras $\mathfrak{L}_4$, $\mathfrak{L}_5$, $\mathfrak{L}_9$, and $\mathfrak{L}_{22}$.

We obtain four different general classes of solutions.

\subsection{Solutions involving only fermionic variables}

For the subalgebra $\mathfrak{L}_1=\{P_1\}$, the reduced system of equations is
\begin{equation}
\begin{split}
P_t=0,\qquad W_t=0,
\end{split}
\label{n1A}
\end{equation}
and the obtained solution is
\begin{equation}
\begin{split}
W=W(\theta_1,\theta_2),\qquad P=P(\theta_1,\theta_2).
\end{split}
\label{n1B}
\end{equation}
This solution has two arbitrary functions of the two fermionic variables $\theta_1$ and $\theta_2$. Using the Taylor expansion of (\ref{n1B}) in $\theta_1$ and $\theta_2$, the solution (\ref{n1B}) can be expressed as
\begin{equation}
\begin{split}
&W(\theta_1,\theta_2)=C_1+\underline{C_2}\theta_1+\underline{C_3}\theta_2+C_4\theta_1\theta_2,\\
& \\
&P(\theta_1,\theta_2)=C_5+\underline{C_6}\theta_1+\underline{C_7}\theta_2+C_8\theta_1\theta_2,
\end{split}
\label{rtie1}
\end{equation}
where $C_1$, $C_4$, $C_5$, $C_8$ are bosonic constants, and $\underline{C_2}$, $\underline{C_3}$, $\underline{C_6}$, $\underline{C_7}$ are fermionic constants.

For the subalgebra $\mathfrak{L}_5=\{P_2\}$, the reduced system of equations is
\begin{equation}
\begin{split}
&P_xW_{\theta_1\theta_2}-\theta_1P_xW_{x\theta_2}+P_{\theta_1\theta_2}W_x-\theta_1P_{x\theta_2}W_x=0,\\
&W_{\theta_1\theta_2}W_x-\theta_1W_{x\theta_2}W_x+(-1)^{\gamma}A(-P_{\theta_1\theta_2})^{\gamma-2}\left[P_{\theta_1}P_{\theta_2}P_{x\theta_1\theta_2}+P_{\theta_1}P_{\theta_2}\theta_1P_{xx\theta_2}+\theta_1P_xP_{\theta_2}P_{x\theta_1\theta_2}\right]\\ &-A(\gamma-2)\theta_1(P_{\theta_1\theta_2})^{\gamma-3}P_{x\theta_2}P_{\theta_1}P_{\theta_2}P_{x\theta_1\theta_2}=0
\end{split}
\label{n5A}
\end{equation}
and the obtained solution in the fermionic sector is
\begin{equation}
W=g((\mu_1-\beta_1)\theta_1+(\mu_2-\beta_2)\theta_2),\qquad
P=f(W)=f((\mu_1-\beta_1)\theta_1+(\mu_2-\beta_2)\theta_2),
\label{n5B}
\end{equation}
This solution has two arbitrary functions, $f$ and $g$, of one fermionic argument $(\mu_1-\beta_1)\theta_1+(\mu_2-\beta_2)\theta_2$. Using the Taylor expansion, the solution (\ref{n5B}) can be reduced to the form
\begin{equation}
W=C_0+\underline{C_1}\left[(\mu_1-\beta_1)\theta_1+(\mu_2-\beta_2)\theta_2\right],\qquad P=C_2+\underline{C_3}\left[(\mu_1-\beta_1)\theta_1+(\mu_2-\beta_2)\theta_2\right],
\end{equation}
where $C_0$ and $C_2$ are bosonic constants and $\underline{C_1}$ and $\underline{C_3}$ are fermionic constants.

For the subalgebra $\mathfrak{L}_9=\{P_1+\varepsilon P_2\}$,
the obtained solution,
\begin{equation}
W=f(-\underline{\alpha_1}\theta_1-\underline{\alpha_2}\theta_2),\qquad
P=g(-\underline{\alpha_1}\theta_1-\underline{\alpha_2}\theta_2),
\label{n9B}
\end{equation}
depends only on the fermionic sector $\theta_1$ and $\theta_2$. This solution has two arbitrary functions, $f$ and $g$, of one bosonic argument $-\underline{\alpha_1}\theta_1-\underline{\alpha_2}\theta_2$. In analogy with the case $\mathfrak{L}_{5}$, this solution can be reduced to
\begin{equation}
W=C_0+\underline{C_1}\left[-\underline{\alpha_1}\theta_1-\underline{\alpha_2}\theta_2\right],\qquad P=C_2+\underline{C_3}\left[-\underline{\alpha_1}\theta_1-\underline{\alpha_2}\theta_2\right],
\end{equation}
where $C_0$ and $C_2$ are bosonic constants and $\underline{C_1}$ and $\underline{C_3}$ are fermionic constants.

For the subalgebra $\mathfrak{L}_{17}=\{P_1+\varepsilon P_2+\underline{\mu}Q_1\}$, 
the obtained solution in the fermionic sector is
\begin{equation}
W=f(\theta_2)+\underline{\mu}(\theta_1),\qquad
P=g(\theta_2)+\underline{\mu}(\theta_1)
\label{n17B}
\end{equation}
which contains two arbitrary functions, $f$ and $g$, of one fermionic argument $\theta_2$. This solution can be written in the form
\begin{equation}
W=C_0+\underline{C_1}\theta_2+\underline{\mu}\theta_1,\qquad P=C_2+\underline{C_3}\theta_2+\underline{\mu}\theta_1,
\end{equation}
where $C_0$ and $C_2$ are bosonic constants and $\underline{C_1}$ and $\underline{C_3}$ are fermionic constants. Since $\underline{\mu}$ is not arbitrary, this solution is not the same as that of $\mathfrak{L}_{5}$.

\subsection{Stationary solutions}

For the subalgebra $\mathfrak{L}_{10}=\{P_1+\underline{\nu}Q_2\}$, 
the obtained stationary solution is
\begin{equation}
W=f(\theta_1)+\underline{\mu}\underline{\eta}(\theta_2-\underline{\nu}x),\qquad
P=g(\theta_1)+\underline{\mu}\underline{\lambda}(\theta_2-\underline{\nu}x)
\label{n10B}
\end{equation}
which contains two arbitrary functions, $f$ and $g$, of two fermionic arguments, $\theta_1$ and $\theta_2-\underline{\nu}x$. This can be reduced to the form
\begin{equation}
W=C_0+\underline{C_1}\theta_1+\underline{\mu}\underline{\eta}(\theta_2-\underline{\nu}x),\qquad P=C_2+\underline{C_3}\theta_1+\underline{\mu}\underline{\lambda}(\theta_2-\underline{\nu}x),
\end{equation}
where $C_0$ and $C_2$ are bosonic constants and $\underline{C_1}$ and $\underline{C_3}$ are fermionic constants. This solution is linear in $x$, $\theta_1$ and $\theta_2$.

For the subalgebra $\mathfrak{L}_{11}=\{P_1+\varepsilon P_2+\underline{\nu}Q_2\}$, 
the obtained stationary solution is
\begin{equation}
W=f(\theta_1)+\underline{\mu}(\theta_2-\underline{\nu}x),\qquad
P=g(\theta_1)+\underline{\lambda}(\theta_2-\underline{\nu}x)
\label{n11B}
\end{equation}
which contains two arbitrary functions, $f$ and $g$, of one fermionic argument $\theta_1$. This can be reduced to the linear form
\begin{equation}
W=C_0+\underline{C_1}\theta_1+\underline{\mu}(\theta_2-\underline{\nu}x),\qquad P=C_2+\underline{C_3}\theta_1+\underline{\lambda}(\theta_2-\underline{\nu}x),
\end{equation}
where $C_0$ and $C_2$ are bosonic constants and $\underline{C_1}$ and $\underline{C_3}$ are fermionic constants.

For the subalgebra $\mathfrak{L}_{13}=\{P_2+\underline{\mu}Q_1\}$, 
the obtained stationary solution is
\begin{equation}
W=f(\theta_2)+\underline{\mu}\eta\theta_1+\underline{\mu}\underline{\nu}x,\qquad
P=g(\theta_2)+\underline{\mu}\lambda\theta_1+\underline{\mu}\underline{\kappa}x
\label{n13B}
\end{equation}
which contains two arbitrary functions, $f$ and $g$, of one fermionic argument $\theta_2$. This can be reduced to the linear form
\begin{equation}
W=C_0+\underline{C_1}\theta_2+\underline{\mu}\underline{\eta}\theta_1+\underline{\mu}\underline{\nu}x,\qquad P=C_2+\underline{C_3}\theta_2+\underline{\mu}\underline{\lambda}\theta_1+\underline{\mu}\underline{\kappa}x,
\end{equation}
where $C_0$ and $C_2$ are bosonic constants and $\underline{C_1}$ and $\underline{C_3}$ are fermionic constants. Since $\underline{\mu}$ is not arbitrary, this solution is not the same as that of $\mathfrak{L}_{11}$.

For the subalgebra $\mathfrak{L}_{21}=\{L_1+\varepsilon P_2\}$, 
the obtained stationary solution is
\begin{equation}
W=x^{3/2}(\underline{\mu}x^{-1/2}\theta_1+\underline{\eta_1}f(\theta_2)),\qquad
P=x^{\left({\gamma+5\over 2\gamma+2}\right)}(\underline{\nu}x^{-1/2}\theta_1+\underline{\eta_2}g(\theta_2))
\label{n21B}
\end{equation}
which contains two arbitrary functions, $f$ and $g$, of one fermionic argument $\theta_2$ and is parametrized by the polytropic exponent $\gamma$. The solution can be expressed as
\begin{equation}
\begin{split}
&W=\underline{\mu}x\theta_1+x^{3/2}\underline{\eta_1}\underline{C_1}\theta_2+x^{3/2}\underline{\eta_1}C_0,\\
& \\
&P=x^{\gamma+5\over 2\gamma+2}\left(\underline{\nu}x^{-1/2}\theta_1+\underline{\eta_2}\underline{C_3}\theta_2+\underline{\eta_2}C_2\right)
\end{split}
\end{equation}
where $C_2$ is a bosonic constant and $\underline{C_1}$ and $\underline{C_3}$ are fermionic constants.

For the subalgebra $\mathfrak{L}_{22}=\{L_1+\underline{\nu}Q_2\}$, 
we obtain the stationary singular solution
\begin{equation}
\begin{split}
&W=\underline{\mu}x\theta_1+x^{3/2}\underline{\eta}\theta_2-\frac{1}{2}x^{3/2}(\ln{x})\underline{\eta}\underline{\nu}\\
&P=\underline{\zeta}x^{\left({4\over 2\gamma+2}\right)}\theta_1+x^{\left({\gamma+5\over 2\gamma+2}\right)}\underline{\rho}\theta_2-\frac{1}{2}x^{\left({\gamma+5\over 2\gamma+2}\right)}(\ln{x})\underline{\rho}\underline{\nu}
\end{split}
\label{n22B}
\end{equation}
which is parametrized by the polytropic exponent $\gamma$.

\subsection{Solutions in the form of a reduced equation}

For the subalgebra $\mathfrak{L}_8=\{L_2\}$, for the case where $\gamma=3$, we obtain the following reduced partial differential equations
\begin{equation}
\begin{split}
{\mathcal A}+\tau_2{\mathcal A}_{\tau_2}-2{\mathcal A}_x{\mathcal A}_{\theta_1\tau_2}+\tau_2{\mathcal A}_x{\mathcal A}_{\theta_1}+2\theta_1{\mathcal A}_x{\mathcal A}_{x\tau_2}-\theta_1\tau_2({\mathcal A}_x)^2=0,
\end{split}
\label{n8A}
\end{equation}
where ${\mathcal A}$ is the superfield described in Table II with invariant $\tau_2=t^{-1/2}\theta_2$. The superfield $W$ is expressed in terms of ${\mathcal A}$ through the relation
\begin{equation}
W=t^{-1/2}{\mathcal A}\left(x,\theta_1,\tau_2\right)
\end{equation}
Here, $P$ is a constant bosonic. This solution represents a non-stationary simple wave involving both bosonic and fermionic variables.

For the subalgebra $\mathfrak{L}_{12}=\{L_2+\varepsilon P_1\}$, for the case where $\gamma=3$, we obtain the following symmetry reduction 
\begin{equation}
\begin{split}
&-{\mathcal A}-\varepsilon {\mathcal A}_{\sigma}-\tau_2{\mathcal A}_{\tau_2}+2{\mathcal A}_{\sigma}{\mathcal A}_{\theta_1\tau_2}-\tau_2{\mathcal A}_{\sigma}{\mathcal A}_{\theta_1}-\varepsilon\tau_2 {\mathcal A}_{\sigma}{\mathcal A}_{\sigma\theta_1}-2\theta_1{\mathcal A}_{\sigma}{\mathcal A}_{\sigma\tau_2}\\&+\theta_1\tau_2({\mathcal A}_{\sigma})^2+\varepsilon\theta_1\tau_2 {\mathcal A}_{\sigma}{\mathcal A}_{\sigma\sigma}=0,
\end{split}
\label{n12A}
\end{equation}
where ${\mathcal A}$ is the superfield described in Table II with invariants $\sigma=x-\frac{1}{2}\varepsilon\ln{t}$ and $\tau_2=t^{-1/2}\theta_2$. The superfield $W$ is expressed in terms of ${\mathcal A}$ through the relation
\begin{equation}
W=t^{-1/2}{\mathcal A}\left(\sigma,\theta_1,\tau_2\right)
\end{equation}
Here, $P$ is a constant bosonic. This solution represents a non-stationary simple wave involving both bosonic and fermionic variables. 

For the subalgebra $\mathfrak{L}_{16}=\{L_2+\underline{\mu}Q_1\}$, for the case where $\gamma=3$, we obtain the following symmetry reduction
\begin{equation}
\begin{split}
&-{\mathcal A}+\underline{\mu}\tau_1{\mathcal A}_{\sigma}-\underline{\mu}{\mathcal A}_{\tau_1}-\tau_2{\mathcal A}_{\tau_2}+2{\mathcal A}_{\sigma}{\mathcal A}_{\tau_1\tau_2}-\tau_2{\mathcal A}_{\sigma}{\mathcal A}_{\tau_1}+\underline{\mu}\tau_2({\mathcal A}_{\sigma})^2+\underline{\mu}\tau_1\tau_2{\mathcal A}_{\sigma}{\mathcal A}_{\sigma\tau_1}\\&-2\tau_1{\mathcal A}_{\sigma}{\mathcal A}_{\sigma\tau_2}+\tau_1\tau_2({\mathcal A}_{\sigma})^2+\underline{\mu}\tau_1\tau_2{\mathcal A}_{\sigma}{\mathcal A}_{\sigma\tau_1}=0
\end{split}
\label{n16A}
\end{equation}
where ${\mathcal A}$ is the superfield described in Table II with invariants $\sigma=x+\frac{1}{2}\underline{\mu}\theta_1\ln{t}$, $\tau_1=\theta_1-\frac{1}{2}\underline{\mu}\ln{t}$ and $\tau_2=t^{-1/2}\theta_2$. The superfield $W$ is expressed in terms of ${\mathcal A}$ through the relation
\begin{equation}
W=t^{-1/2}{\mathcal A}\left(\sigma,\tau_1,\tau_2\right)
\end{equation}
Here, $P$ is a constant bosonic. This solution represents a non-stationary simple wave involving both bosonic and fermionic variables.

For the subalgebra $\mathfrak{L}_{20}=\{L_2+\varepsilon P_1+\underline{\mu}Q_1\}$, for the case where $\gamma=3$, we obtain the following symmetry reduction
\begin{equation}
\begin{split}
&-{\mathcal A}+\underline{\mu}\tau_1{\mathcal A}_{\sigma}-\varepsilon {\mathcal A}_{\sigma}-\underline{\mu}{\mathcal A}_{\tau_1}-\tau_2{\mathcal A}_{\tau_2}+2{\mathcal A}_{\sigma}{\mathcal A}_{\tau_1\tau_2}-\tau_2{\mathcal A}_{\sigma}{\mathcal A}_{\tau_1}+\underline{\mu}\tau_2({\mathcal A}_{\sigma})^2\\&+\underline{\mu}\tau_1\tau_2{\mathcal A}_{\sigma}{\mathcal A}_{\sigma\tau_1}-\varepsilon\tau_2{\mathcal A}_{\sigma}{\mathcal A}_{\sigma\tau_1}-2\tau_1{\mathcal A}_{\sigma}{\mathcal A}_{\sigma\tau_2}+\tau_1\tau_2({\mathcal A}_{\sigma})^2+\varepsilon\tau_1\tau_2{\mathcal A}_{\sigma}{\mathcal A}_{\sigma\sigma}\\&+\underline{\mu}\tau_1\tau_2{\mathcal A}_{\sigma}{\mathcal A}_{\sigma\tau_1}=0
\end{split}
\label{n20A}
\end{equation}
where ${\mathcal A}$ is the superfield described in Table II with invariants $\sigma=x+\frac{1}{2}\underline{\mu}\theta_1\ln{t}-\frac{1}{2}\varepsilon\ln{t}$, $\tau_1=\theta_1-\frac{1}{2}\underline{\mu}\ln{t}$ and $\tau_2=t^{-1/2}\theta_2$. The superfield $W$ is expressed in terms of ${\mathcal A}$ through the relation
\begin{equation}
W=t^{-1/2}{\mathcal A}(\sigma,\tau_1,\tau_2)
\end{equation}
Here, $P$ is a constant bosonic. This solution represents a non-stationary simple wave involving both bosonic and fermionic variables.

For the subalgebra $\mathfrak{L}_{24}=\{L_1+kL_2\}$, for the case where $\gamma=2$ and $k=7$, we obtain the following symmetry reduction
\begin{equation}
\begin{split}
&-4{\mathcal A}-2\sigma {\mathcal A}_{\sigma}-\tau_1{\mathcal A}_{\tau_1}-7\tau_2{\mathcal A}_{\tau_2}+14{\mathcal A}_{\tau_1\tau_2}{\mathcal A}_{\sigma}-4\tau_2{\mathcal A}_{\tau_1}{\mathcal A}_{\sigma}-2\sigma\tau_2{\mathcal A}_{\sigma\tau_1}{\mathcal A}_{\sigma}-\tau_2{\mathcal A}_{\tau_1}{\mathcal A}_{\sigma}\\ &-14\tau_1{\mathcal A}_{\sigma\tau_2}{\mathcal A}_{\sigma}+4\tau_1\tau_2({\mathcal A}_{\sigma})^2+2\sigma\tau_1\tau_2{\mathcal A}_{\sigma\sigma}{\mathcal A}_{\sigma}=0
\end{split}
\label{n24A}
\end{equation}
where ${\mathcal A}$ is the superfield described in Table II with invariants $\sigma=t^{-\frac{1}{7}}x$, $\tau_1=t^{-\frac{1}{14}}\theta_1$ and $\tau_2=t^{-\frac{1}{2}}\theta_2$. The superfield $W$ is expressed in terms of ${\mathcal A}$ through the relation
\begin{equation}
W=t^{-\frac{2}{7}}{\mathcal A}\left(\sigma,\tau_1,\tau_2\right)
\end{equation}
Here, $P$ is a constant bosonic. This solution represents a non-stationary simple wave involving both bosonic and fermionic variables.

\subsection{Wave solutions}

For the subalgebra $\mathfrak{L}_3=\{P_1+\underline{\mu}Q_1\}$, the reduced system of equations is
\begin{equation}
\begin{split}
&P_t-2\underline{\mu}P_{\tau_1}P_{\tau_1\theta_2}-2\underline{\mu}P_{\tau_1}\theta_2P_{t\tau_1}+P_{\tau_1\theta_2}W_x+\theta_2P_{t\tau_1}W_x+\underline{\mu}\tau_1P_{\tau_1\theta_2}W_x+\underline{\mu}\tau_1\theta_2P_{t\tau_1}W_x=0,\\
&W_t+2W_{\tau_1\theta_2}W_x+2\theta_2W_{t\tau_1}W_x+2\underline{\mu}\tau_1W_{\tau_1\theta_2}W_x+2\underline{\mu}\tau_1\theta_2W_{t\tau_1}W_x=0
\end{split}
\label{n3A}
\end{equation}
and the obtained solution is
\begin{equation}
\begin{split}
W=\underline{\mu}tf_1(\tau_1,\tau_2)+f_2(\tau_1,\tau_2),\qquad P=\underline{\mu}tg_1(\tau_1,\tau_2)+g_2(\tau_1,\tau_2),
\end{split}
\label{n3B}
\end{equation}
where $f_1$ and $g_1$ are fermionic-valued functions and $f_2$ and $g_2$ are bosonic-valued functions of the two arguments
\begin{equation}
\tau_1=\theta_1-\underline{\mu}x\mbox{ and }\tau_2=\theta_2+\underline{\mu}t.
\end{equation}
The functions $f_1$, $f_2$, $g_1$ and $g_2$ satisfy the relations
\begin{equation}
\begin{split}
f_1=2f_{2,\tau_1\tau_2}f_{2,\tau_1}-f_{2,\tau_2},\qquad g_1=g_{2,\tau_1\tau_2}f_{2,\tau_1}+2g_{2,\tau_1}g_{2,\tau_1\tau_2}-g_{2,\tau_2},
\end{split}
\label{n3C}
\end{equation}
where $f_2$ and $g_2$ are completely arbitrary functions of $\tau_1$ and $\tau_2$. This solution has a degree of freedom of two arbitrary functions of two variables which combine the fermionic variables $\theta_1$ and $\theta_2$ with the bosonic variables $x$ and $t$. It represents a propagation wave in both the bosonic and fermionic variables. Since the rank of the matrix
\begin{equation}
\begin{bmatrix} {\partial W\over \partial x} & {\partial W\over \partial t} & {\partial W\over \partial \theta_1} & {\partial W\over \partial \theta_2} \\ & & & \\ {\partial P\over \partial x} & {\partial P\over \partial t} & {\partial P\over \partial \theta_1} & {\partial P\over \partial \theta_2} \end{bmatrix}
\end{equation}
is $2$, this solution is called a Riemann scattering double wave. For the specific case where we choose
\begin{equation}
f_2=e^{k\tau_1\tau_2},\qquad g_2=e^{l\tau_1\tau_2},
\end{equation}
we obtain the propagating wave solution
\begin{equation}
\begin{split}
&W=\underline{\mu}t\left(2k^2\theta_2+k\theta_1\right)+1+k\theta_1\underline{\mu}t+k\theta_2\underline{\mu}x+k\theta_1\theta_2,\\
& \\
&P=\underline{\mu}t\left(kl+2l^2\right)\theta_2+\underline{\mu}tl\theta_1+1+l\theta_1\underline{\mu}t+l\theta_2\underline{\mu}x+l\theta_1\theta_2
\end{split}
\end{equation}

For the subalgebra $\mathfrak{L}_4=\{L_1\}$,
we have obtained the explicit stationary solution in monomial form
\begin{equation}
\begin{split}
W=x^{3/2}\underline{\mu}\theta_2,\qquad P=x^{\left({\gamma+5\over 2\gamma+2}\right)}\underline{\nu}\theta_2
\end{split}
\label{n4B}
\end{equation}
which depends on the polytropic exponent $\gamma$.

For the subalgebra $\mathfrak{L}_7=\{P_2+\underline{\nu}Q_2\}$,
the obtained non-stationary simple wave solution involving both bosonic and fermionic variables is
\begin{equation}
W=g(\underline{\mu}t+k\theta_1+\underline{\mu}\underline{\nu}\theta_2),\qquad
P=f(\underline{\mu}t+k\theta_1+\underline{\mu}\underline{\nu}\theta_2)
\label{n7B}
\end{equation}
which involves a bosonic constant $k$ and two arbitrary functions, $f$ and $g$, of one fermionic argument $\underline{\mu}t+k\theta_1+\underline{\mu}\underline{\nu}\theta_2$.

For the subalgebra $\mathfrak{L}_{23}=\{L_1+\varepsilon P_2+\underline{\nu}Q_2\}$, 
the obtained solution is the simple wave
\begin{equation}
W=\underline{\mu}x^{3/2}A(\sigma)\qquad
P=\underline{\eta}x^{\left({\gamma+5\over 2\gamma+2}\right)}\underline{\alpha}
\label{n23B}
\end{equation}
which contains one arbitrary function $A$ of one argument $\sigma=t+\frac{1}{2}\underline{\nu}\theta_2\ln{x}-\frac{1}{2}\varepsilon\ln{x}$ and is parametrized by the polytropic exponent $\gamma$. In the specific case where one chooses $A(\sigma)=e^{-k\sigma}$, we obtain
\begin{equation}
W=\underline{\mu}x^{3-\varepsilon\over 2}e^{-kt}\left(1+\frac{1}{2}\underline{\nu}\theta_2\ln{x}\right),\qquad P=\underline{\eta}x^{\gamma+5\over 2\gamma+2}\underline{\alpha}.
\end{equation}
Here, $W$ involves damping in time $t$.

The remaining subalgebras have invariants which possess a non--standard structure in the sense that one does not obtain standard symmetry reductions from them (as described in \cite{SSG}). Such subalgebras are distinguished by the fact that each of them admits an invariant expressed in terms of an arbitrary function of the superspace variables, multiplied by a fermionic constant.
We list in Table III the invariants expressed in terms of an arbitrary function of the superspace variables for each case.\\\\

\begin{table}[htbp]
  \begin{center}
\caption{List of subalgebras with nonstandard invariants. In each case, $f$ is an arbitrary function of its arguments.}
\vspace{3mm}
\setlength{\extrarowheight}{4pt}
\begin{tabular}{|c|c|}\hline
Subalgebra & Non-standard invariant \\[0.5ex]\hline\hline
$\mathfrak{L}_2=\{\underline{\mu}Q_1\}$ & $\underline{\mu}f\left(x,t,\theta_1,\theta_2,W,P\right)$ \\\hline
$\mathfrak{L}_6=\{\underline{\nu}Q_2\}$ & $\underline{\nu}f\left(x,t,\theta_1,\theta_2,W,P\right)$ \\\hline
$\mathfrak{L}_{14}=\{\underline{\mu}Q_1+\underline{\nu}Q_2\}$ & $\underline{\mu}\underline{\nu}f\left(x,t,\theta_1,\theta_2,W,P\right)$ \\\hline
$\mathfrak{L}_{15}=\{P_2+\underline{\mu}Q_1+\underline{\nu}Q_2\}$ & $\underline{\mu}\underline{\nu}f\left(x,\theta_1,\theta_2,W,P\right)$ \\\hline
$\mathfrak{L}_{18}=\{P_1+\underline{\mu}Q_1+\underline{\nu}Q_2\}$ & $\underline{\mu}\underline{\nu}f\left(t,\theta_1,\theta_2,W,P\right)$ \\\hline
$\mathfrak{L}_{19}=\{P_1+\varepsilon P_2+\underline{\mu}Q_1+\underline{\nu}Q_2\}$ & $\underline{\mu}\underline{\nu}f\left(\theta_1,\theta_2,W,P\right)$ \\\hline
\end{tabular}
  \end{center}
\end{table}

\section{Concluding remarks and future outlook}

The objective of this paper was to construct a supersymmetric extension of the polytropic gas dynamics equations through a superspace involving two independent variables and two bosonic superfields and to perform a systematic group-theoretical analysis of this extension. We found a Lie superalgebra of symmetries of the proposed supersymmetric model (\ref{c1}) which included two translations, two dilations and two supersymmetric transformations. Decomposing this superalgebra into a direct sum, we used the Goursat method to classify its one-dimensional subalgebras into 24 conjugacy classes under the action of the associated supergroup. By using the symmetry reduction method, we determined the invariant solutions of the supersymmetric model under consideration. By postulating a wave-like form for the solutions, we obtained a number of classes of solutions involving arbitrary functions of one or two variables. In the particular case where the arbitrary functions involved only $\theta_1$ and $\theta_2$, we were able to explicitly find the most general form of the solution through a truncated Taylor expansion. We constructed algebraic-type solutions and propagation waves (including simple and double waves) in explicit form. We conclude that the solutions of the supersymmetric polytropic gas dynamics model (\ref{c1}) include propagation waves and their superpositions involving both bosonic and fermionic variables. This phenomenon is analogous to that of the multisolitonic wave solutions obtained in the case of supersymmetric extensions of integrable models.

The following question can be considered. The freedom of the obtained solution can be found from the Cauchy data when $t=0$. However, the aspect of the Cauchy problem involving uniqueness and continuous dependence on initial conditions remains an open problem for the case of hydrodynamic-type equations. This will be addressed in a future work.

\subsection*{Acknowledgements}

The authors thank Professor R. Conte (\'{E}cole Normale Sup\'{e}rieure, CMLA, Cachan and Centre d'\'{E}nergie Atomique de Saclay) for helpful discussions on this topic. This work was supported by research grants from NSERC of Canada.

{}

\label{lastpage}

\begin{thebibliography}{99}
\bibliographystyle{plain}

\bibitem{Jackiw}
R. Jackiw, \em{A Particle Theorist's Lectures on Supersymmetric
Non-Abelian Fluid Mechanics and d-branes}\em\ (Springer-Verlag, New
York, 2002).

\bibitem{Das}
A. Das and Z. Popowicz, %``Supersymmetric polytropic gas dynamics,''
\em{Phys. Lett. A}\em\ \textbf{296}, 15 (2002). %15--26.

\bibitem{Bergner}
Y. Bergner and R. Jackiw, %``Integrable supersymmetric fluid mechanics from superstrings,''
\em{Phys. Lett. A}\em\ \textbf{284}, 146 %-151 
(2001).

\bibitem{Polychronakos}
R. Jackiw and A. P. Polychronakos, %``Supersymmetric Fluid Mechanics,''
\em{Phys. Rev. D}\em\ \textbf{62}, 085019 (2000).

\bibitem{Roz}
B. L. Rozdestvenskii and N. N. Janenko, \em{Systems of Quasilinear Equations and Their Applications to Gas Dynamics}\em, vol. 55 (American Mathematical Society, Providende, R.I., 1983).

\bibitem{Hariton9}
A. M. Grundland and A. J. Hariton, %``Supersymmetric version of a hydrodynamic system in Riemann invariants and its solutions,''
\em{J. Math. Phys.}\em\ \textbf{49},4, 043502
(2008).

\bibitem{Hariton10}
A. M. Grundland and A. J. Hariton, %``N = 2 supersymmetric extension of a hydrodynamic system in Riemann invariants,''
\em{J. Math. Phys.}\em\ \textbf{50},4, 073508
(2009).

\bibitem{Ablowitz}
M. J. Ablowitz, P. A. Clarkson, \em{Solitons, Nonlinear Evolution Equations and Inverse Scattering}\em\ (Cambridge University Press, Cambridge, 1991).

\bibitem{Rogers}
C. Rogers and W. K. Schief, \em{B\"{a}cklund and Darboux Transformations}\em\ (Cambridge University Press, Cambridge, 2002).

\bibitem{Mathieu}
P. Mathieu, %``Supersymmetric extension of the Korteweg--de Vries equation,''
\em{J. Math. Phys.}\em\ \textbf{29}, 2499 (1988). %2499--2506.

\bibitem{Labelle}
P. Labelle, P. Mathieu, %``A new N=2 supersymmetric Korteweg--de Vries equation,''
\em{J. Math. Phys.}\em\ \textbf{32}, 923 (1991). %923--927.

\bibitem{Manin}
Y. I. Manin and A. O. Radul, %``A supersymmetric extension of the Kadomtsev-Petviashvili hierarchy,''
\em{Commun. Math. Phys.}\em\ \textbf{98}, 65 (1985). %65--77.

\bibitem{Hariton4}
A. J. Hariton, %``Supersymmetric extension of the scalar Born-Infeld equation,''
\em{J. Phys. A: Math. Gen.}\em\ \textbf{39}, 7105 %--7114
(2006).

\bibitem{Hariton8}
A. M. Grundland and A. J. Hariton, %``A supersymmetric version of a Gaussian irrotational compressible fluid flow,''
\em{J. Phys. A: Math. Theor.}\em\ \textbf{40}, 15113 %--15129
(2007).

\bibitem{Fatyga}
B. W. Fatyga, V. A. Kostelecky and D. R. Truax, %``Grassmann‐valued fluid dynamics,''
\em{J. Math. Phys.}\em\ \textbf{30}, 1464 (1989). %1464--1472.

\bibitem{Grammaticos}
B. Grammaticos, A. Ramani, A. S. Carstea, %``Bilinearization and soliton solutions of the $N = 1$ supersymmetric sine--Gordon equation,''
\em{J. Phys. A: Math. Gen.}\em\ \textbf{34}, 4881 (2001). %4881--4886.

\bibitem{Siddiq1}
M. Siddiq, M. Hassan, %``On the linearization of the super sine--Gordon equation,''
\em{Europhys. Lett.}\em\ \textbf{70}, 149 (2005). %149--154.

\bibitem{Siddiq2}
M. Siddiq, M. Hassan, U. Saleem, %``On Darboux transformation of the supersymmetric sine--Gordon equation,''
\em{J. Phys. A}\em\ \textbf{39}, 7313 (2006). %7313--7318.

\bibitem{Chaichian}
M. Chaichian, P. P. Kulish, %``On the method of inverse scattering problem and B\"{a}cklund transformations for supersymmetric equations,''
\em{Phys. Lett. B}\em\ \textbf{78}, 413 (1978). %413--416.

\bibitem{Liu3}
Q. P. Liu, M. Manas, %``Pfaffian solutions for the Manin--Radul--Mathieu SUSY KdV and SUSY sine--Gordon equations,''
\em{Phys. Lett. B}\em\ \textbf{436}, 306 (1998). %306--310.

\bibitem{Peradzynski}
Z. Peradzynski, ``Geometry of interactions of riemann waves,'' in \em{Advances in Nonlinear Waves (Research Notes in Math 111, vol. 3}\em, Ed. L. Debnath (Boston, Pitman, 1985), 244. %--285.

\bibitem{Cornwell}
J. F. Cornwell, \em{Group Theory in Physics, Volume 3}\em\ (Academic Press, London, 1989).

\bibitem{DeWitt}
B. DeWitt, \em{Supermanifolds}\em\ (Cambridge University Press, Cambridge, 1984).

\bibitem{Olver}
P. J. Olver, \em{Applications of Lie Groups to Differential Equations}\em\ (Springer-Verlag, New York, 1986).

\bibitem{SSG}
A. M. Grundland, A. J. Hariton, L. \v{S}nobl, %``Invariant solutions of the supersymmetric sine--Gordon equation,''
\em{J. Phys. A: Math. Theor.}\em\ \textbf{42}, 335203 (2009). %335203.

\bibitem{ARG}
A. Rogers, %``Super Lie groups: global topology and local structure,''
\em{J. Math. Phys.}\em\ \textbf{22}, 939 (1981). %939.

\bibitem{ARG2}
A. Rogers, %``A global theory of supermanifolds,''
\em{J. Math. Phys.}\em\ \textbf{21}, 1352 (1980). %1352.

\bibitem{Roelofs2}
G. H. M. Roelofs and N. W. van den Hijligenberg, %``Prolongation structures for supersymmetric equations,''
\em{J. Phys. A: Math. Gen.}\em\ \textbf{23}, 5117 (1990). %5117--5130.

\bibitem{Berezin}
F. A. Berezin, \em{The Method of Second Quantization}\em\ (Academic Press, New York, 1966).

\bibitem{Patera}
J. Patera, P. Winternitz, R. T. Sharp and H. Zassenhaus, %``Continuous subgroups of the fundamental groups of physics,''
\em{J. Math. Phys.}\em\ \textbf{18}, 2259 (1977). %2259.

\bibitem{Winternitz1}
P. Winternitz, ``Lie Groups and Solutions of Nonlinear Partial
Differential Equations,'' in \em{Integrable Systems, Quantum Groups and Quantum
Field Theories}\em, Eds. L.A. Ibort and M.A. Rodriguez (Kluwer, Dordrech, 1993), 429.

\bibitem{Winternitz2}
P. Winternitz, ``Group Theory and Exact Solutions of Partially Integrable Differential Systems,'' in \em{Partially Integrable Evolution Equations in Physics}\em, Eds. R. Conte and N. Boccara (Kluwer, Netherlands, 1990), 515. %--567.

\bibitem{Kac}
V. Kac, ``Classification of supersymmetries,'' in
\em{Proceedings of the ICM, Vol. 1}\em\ (Beijing) (2002), 319.

\bibitem{Goursat}
E. Goursat, 1889 %``Sur les substitutions orthogonales,''
\em{Ann. Sci. Ec. Normale Sup.}\em \textbf{3}, 6, 9 %--102 
(1889).

\bibitem{DuVal}
P. DuVal, \em{Homographies, Quaternions and Rotations}\em\ (Clarendon Press, Oxford, 1964).

\bibitem{Clarkson}
P. A. Clarkson and P. Winternitz, ``Symmetry reduction and exact solutions of nonlinear partial differential equations,'' in
 \em{The Painlev\'{e} Property, One Century Later}\em, Ed. R. Conte (Springer-Verlag, New-York, 1999), 597. %-669.



\end{thebibliography}
\end{document}